\documentclass[journal]{IEEEtran}

\usepackage{xcolor}
\usepackage{booktabs}
\usepackage{subcaption}

\usepackage{placeins}

\usepackage{multirow}
\usepackage{booktabs}

\usepackage{longtable}
\usepackage{array}

\newcommand{\best}[1]{\boldmath\textbf{#1}\unboldmath}

\usepackage[nocompress]{cite}

\ifCLASSINFOpdf
  \usepackage[pdftex]{graphicx}
 
\else
  
\fi

\usepackage{amsmath,amssymb}

\definecolor{revteal}{RGB}{31, 89, 140}
\newcommand{\zx}[1]{\textcolor{revteal}{#1}}
\newcommand{\supp}[1]{\textcolor{red}{#1}}

\renewcommand{\supp}[1]{#1}
\renewcommand{\zx}[1]{#1}

\begin{document}

\title{Self-Supervised Multibeam Sonar Restoration via Coherent Forward Modeling}

\author{Zixin~Wang,
        James~Hopgood,
        and~Mike~Davies%
\thanks{The authors are with the School of Engineering, University of Edinburgh, Edinburgh EH9 3FG, U.K. (e-mail: Z.Wang-421@sms.ed.ac.uk; James.Hopgood@ed.ac.uk; Mike.Davies@ed.ac.uk).}}

\maketitle

\bstctlcite{IEEEexample:BSTcontrol}

\begin{abstract}
Multibeam imaging sonar produces intensity images degraded by coherent speckle and system-induced blur, while clean reflectivity targets are rarely available. Existing despeckling methods often rely on simplified intensity-domain noise models, synthetic clean targets, or blind-spot self-supervision, and typically do not jointly model deblurring and speckle suppression.

We propose a physics-consistent self-supervised framework for joint despeckling and deblurring of multibeam imaging sonar. A stochastic coherent image-formation model yields a conditional-mean forward operator that relates latent scene reflectivity to expected intensity through the squared magnitude of the system point spread function.
We then use registered multi-frame sonar sequences to train without clean ground truth images, showing that, under an idealized inter-frame conditional-mean condition, the expected paired loss equals the corresponding forward-prediction loss up to a network-independent constant.
To stabilize the ill-conditioned inverse problem, we further adapt robust equivariant imaging using stochastic coherent simulation and introduce a paired forward prediction error metric for ground-truth-free evaluation. Experiments on two real multibeam sonar datasets demonstrate a favourable balance between forward-prediction agreement, homogeneous-region speckle suppression, and preservation of visible structural boundaries compared with representative baselines.

\end{abstract}


\IEEEpeerreviewmaketitle

\section{Introduction}
\IEEEPARstart{M}{ultibeam} imaging sonar is widely used for underwater perception, mapping, inspection, and robotic navigation because it remains effective in turbid water and low-visibility environments \cite{EPI_Metrics, sonar_background, sonar_challenge}. However, sonar images are often degraded by coherent speckle and system-induced blur, which obscure small targets, weaken object boundaries, and reduce the reliability of downstream tasks such as detection, segmentation, and scene interpretation \cite{underwater_detection, underwater_segmentation, scene_interpretation}. Effective restoration therefore requires not only speckle suppression, but also recovery of structurally faithful and spatially sharp image content.

A central difficulty is that practical imaging sonars usually provide detected intensity images, while the dominant degradations arise from coherent acoustic image formation. The beamformed complex field is formed by the coherent superposition of echoes from unresolved scatterers and is subsequently converted to intensity. This process gives rise to signal-dependent, non-Gaussian speckle rather than simple additive noise \cite{array_signal_processing, goodman_fundamental_property, sonar_despeckling_bg}. In addition, the system response acts on the coherent field before intensity detection, so speckle correlation and blur are coupled through the effective point spread function (PSF) of the imaging system \cite{speckle_gamma_book, sonar_principles, goodman_speckle_book}. Since clean reflectivity ground truth is rarely available, imaging-sonar restoration must address a coupled inverse problem without ground truth: joint despeckling and deblurring from intensity-only observations.

Existing sonar despeckling methods, together with the broader Synthetic Aperture Radar (SAR) despeckling literature, are often based on simplified intensity-domain statistical models. Classical approaches commonly treat speckle as multiplicative noise, frequently with exponential or Gamma statistics, and sometimes transform it into an additive form in the logarithmic domain \cite{Frost1982, speckle_gamma_book, despeckle_tutorial_SAR, log_despeckle}. These assumptions have motivated local adaptive filters, Bayesian estimators, diffusion-based methods, and nonlocal patch aggregation approaches \cite{Despeckle_SAR_filter_local_statistic, despeckle_adaptive_smoothing_filter_SAR, PPB, SAR_BM3D}. However, these formulations primarily describe marginal or locally filtered intensity statistics. Spatial correlation is typically introduced through filtering windows, patch matching, aggregation, or hand-specified correlation structures \cite{despeckle_correlation_filter}.
Consequently, they do not provide a sonar-specific analytical observation model in which spatially correlated speckle and point spread function (PSF) blur are generated jointly.

Learning-based despeckling methods can achieve strong empirical performance, but supervised training generally requires clean reflectivity targets, which are rarely obtainable in real sonar imaging \cite{despeckle_Blind_reduction, additive_speckle_learning, SimNFND_speckle_supervised, log_despeckle, SIDTGAN_supervised_despeckle}. Many methods therefore rely on synthetic clean images, simulated speckle corruption, or architectures transferred from optical denoising \cite{multivariate, optical_despeckle}. These strategies alleviate the lack of labeled sonar data, but they can introduce a mismatch between the training degradation and real sonar observations, where coherent speckle, spatial correlation, system blur, platform motion, and sensor-specific effects are coupled by the acquisition process.
A recent despeckling method adopts a multiplicative \(y=xn\) model with Gamma-distributed speckle and constrains log-domain residual statistics \cite{PhysicsGuided}. However, its formulation remains a single-image despeckling framework and does not explicitly model PSF-coupled coherent image formation or the spatially correlated speckle induced by the imaging process.

Self-supervised methods reduce the dependence on clean targets by constructing supervision from noisy measurements themselves. Blind-spot, masking-based, and subsampling-based approaches prevent the network from directly observing the prediction target and have been widely used for image despeckling \cite{Speckle2Void, noise2void, noise2self, neighbor2neighbor, APBSN, LG_BPN, SS_BSN, Blind2unblind, SEGSID}.
\zx{These methods usually rely on a conditional-independence assumption: the corruption at the prediction target should be independent of the visible measurements given the underlying clean signal. This assumption is difficult to satisfy in imaging sonar because coherent propagation and the system response introduce spatially correlated speckle. Enlarging the blind spot or excluding neighbouring pixels can reduce noise leakage, but it also removes local spatial context that is important for preserving structure and undoing system blur.}

\zx{Noisy-pair learning provides a complementary route to self-supervision. Noise2Noise showed that clean targets can be replaced by independently corrupted observations with the same conditional mean, and Frame2Frame extended this idea to registered neighbouring video frames \cite{noise2noise, frame2frame}. Related principles have also been explored for multitemporal SAR despeckling and registered sonar sequences \cite{SAR2SAR, RegN2N}. In coherent sonar, however, neighbouring frames are not simply independent additive-noise observations: they are affected by viewpoint change, platform motion, registration error, coherent speckle decorrelation, and the system point spread function. Therefore, a registered-frame objective must specify not only how frames are paired, but also which physical domain the network output represents.}

\zx{Our preliminary work, PI-SSD \cite{RegN2N}, used registered sonar frame pairs for self-supervised despeckling. However, it matches the network output directly to the second observation, \(\|f_\theta(y_1)-y_2\|_2^2\), without applying the PSF-dependent forward operator to the output. The network therefore targets a still PSF-blurred intensity image.}

\zx{Building on PI-SSD, the present work uses the coherent image-formation model explicitly in the learning objective to formulate a self-supervised inverse problem, rather than treating paired frames as generic noisy images. We derive the conditional mean intensity learning goal for the coherent speckle-contaminated forward model, showing how the expected intensity observation is related to the scene reflectivity and the squared magnitude of the system PSF. This provides a physically meaningful forward prediction target for training from registered multi-frame sonar observations without clean reflectivity ground truth.}
\zx{The resulting inverse problem remains ill-conditioned, since different reflectivity estimates can produce similar intensity observations. We therefore extend robust equivariant imaging (REI) \cite{chen2021equivariant, robustEI} to sonar restoration by adapting it to the derived conditional mean learning goal and incorporating it within a Noise2Noise-type self-supervised training framework. This combines a physics-consistent paired loss, which exploits registered sonar frames with approximately decorrelated speckle, with REI regularization to constrain the reconstruction under known transformations. Finally, using the same paired-frame protocol, we introduce paired forward prediction error (PFPE) to assess ground-truth-free forward-prediction consistency with registered paired observations.}

This work treats multibeam-sonar restoration as a physics-consistent self-supervised inverse problem, rather than as generic image denoising. The network estimates a latent reflectivity image, and supervision is imposed only after projection through a coherent conditional-mean forward model. This distinguishes the proposed framework from blind-spot denoising and direct noisy-pair despeckling methods, which operate primarily in the observation/image domain.
The main contributions of this work are summarized as follows:
\begin{itemize}
    \item We develop a coherent-formation-based statistical observation model for imaging sonar restoration. The model explains spatially correlated speckle and system-induced blur as coupled consequences of coherent propagation and the sonar PSF, rather than imposing speckle correlation through intensity-domain post-filtering.

    \item \zx{We derive the corresponding conditional mean intensity learning goal for the coherent speckle-contaminated forward model. This provides a physics-consistent forward prediction objective for learning from detected intensity measurements without requiring clean reflectivity ground truth.}

    \item \zx{We propose a self-supervised reconstruction framework for joint despeckling and deblurring using registered multi-frame sonar observations. Unlike direct noisy-pair matching, the paired loss constrains the forward projection $A f_\theta(y_1)$ of the latent reconstruction, while REI regularization uses stochastic coherent simulation from the same image-formation model to reduce the ill-conditioning of sonar intensity reconstruction.}

    \item \zx{Finally, based on the same multi-frame analysis, we introduce paired forward prediction error (PFPE) as a ground-truth-free measure of measurement-domain forward-prediction consistency under a fixed paired-frame protocol. PFPE provides a complementary evaluation criterion for real sonar data where clean reflectivity ground truth is unavailable.}
    
\end{itemize}

\section{Physical Observation Model}
\label{sec:physical_model}
\zx{Multibeam imaging sonar systems provide detected intensity images, but these images are generally impacted by speckle and blur as a byproduct of the coherent imaging process. Let $\Omega$ denote the discrete range-angle image grid of the sonar, with pixel index $i\in\Omega$. We write $x\in\mathbb{R}_{+}^{|\Omega|}$ for the latent scene reflectivity in intensity units, where $x_i\geq 0$ is the reflectivity at pixel $i$. An observed intensity image is denoted by $y\in\mathbb{R}_{+}^{|\Omega|}$.}

We model one observed intensity realization as
\begin{equation}
\label{eq:forward_model}
y=\left|u*h\right|^2=
\left|(\sqrt{x}\cdot g)*h\right|^2,
\end{equation}
where $u=\sqrt{x}\cdot g\in\mathbb{C}^{|\Omega|}$ is the speckle-modulated coherent field, $\cdot$ denotes pointwise multiplication, and $*$ denotes spatial convolution. The magnitude square in \eqref{eq:forward_model} is applied pointwise.

Under the fully developed speckle model \cite{goodman_speckle_book}, $g\in\mathbb{C}^{|\Omega|}$ is a circular complex Gaussian speckle field. At each range-angle bin $i\in\Omega$,
\begin{equation}
\label{eq:speckle_field}
g_i\sim\mathcal{CN}(0,1),
\qquad
\mathbb{E}[|g_i|^2]=1.
\end{equation}
\zx{Before processing by the imaging system, this gives $\mathbb{E}[|u_i|^2\mid x]=x_i$.}

\zx{Here, $h\in\mathbb{C}^{|\Omega|}$ is the coherent point spread function (PSF) of the sonar system, written as}
\begin{equation}
\label{eq:psf_decomposition}
h := h(t,\vartheta)=R(t)B(\vartheta),
\end{equation}
where $t$ denotes two-way travel time, $R(t)$ is the temporal response, and $B(\vartheta)$ is the angular beam pattern \cite{DASbeamforming, sonar_principles, sonar_systems}. For simplicity, we use a local shift-invariant approximation of $h$ evaluated around the central beam direction, which keeps the forward operator tractable for local image regions.

\begin{figure}[t]
    \centering
    \includegraphics[width=0.88\columnwidth]{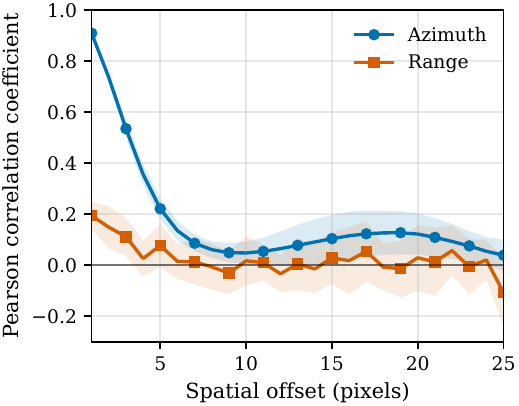}
\caption{\zx{
Empirical spatial correlation of detected sonar intensity measured from homogeneous ROIs in a representative Boat frame. Correlation is computed along the range and azimuth directions; offsets start from one pixel to exclude the trivial zero-offset self-correlation. Curves show the mean across ROIs, with shaded bands indicating one standard deviation.
}}
    \label{fig:spatial_speckle_correlation}
\end{figure}

In imaging sonar, the coherent PSF mixes neighbouring coherent returns before magnitude-square detection. Under fully developed speckle, the single-look intensity has exponential marginal statistics \cite{goodman_fundamental_property, goodman_speckle_book}. Since the coherent field is filtered before detection, the resulting intensity samples are spatially correlated rather than independent pointwise speckle. \zx{This spatial correlation is illustrated in Fig.~\ref{fig:spatial_speckle_correlation}, where empirical correlation decays with range and azimuth spatial offset. This also explains a limitation of blind-spot self-supervision for sonar restoration: reducing leakage from spatially correlated speckle requires masking a local neighbourhood around the prediction target, but this removes precisely the nearby structural information that is important for joint despeckling and deblurring.}

\section{Methodology}

\begin{figure*}[t]
  \centering
  \includegraphics[width=\textwidth]{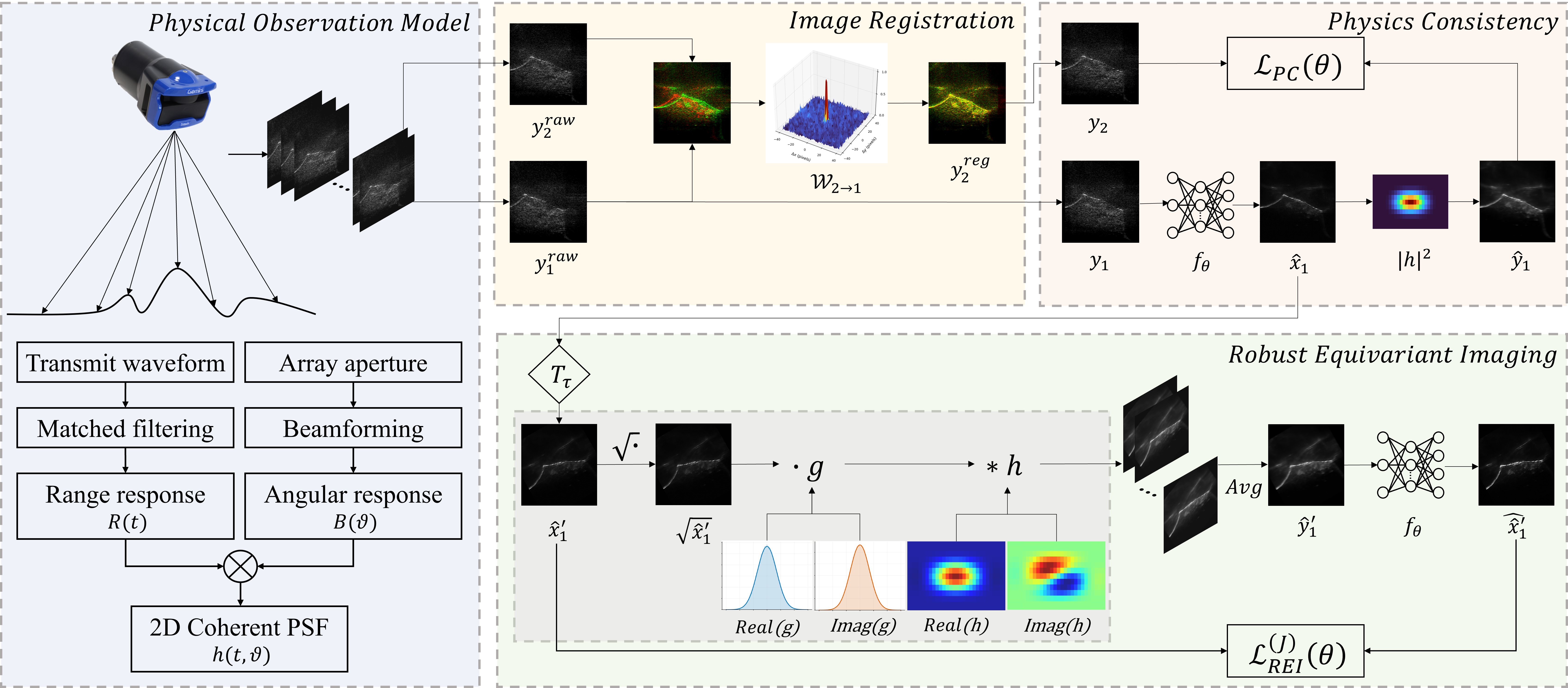}
  \caption{
Overview of the proposed framework. 
The framework contains four components: a coherent physical observation model, an image-registration module that constructs registered paired observations, a physics-consistency branch that compares the forward projection of the reconstruction with a registered frame, and a robust equivariant imaging branch that regularizes the latent reconstruction. 
At inference time, only the trained reconstruction network is used and it operates on a single sonar image.
}
  \label{fig:framework_MCEI}
\end{figure*}

\subsection{Problem Formulation}

Given detected sonar intensity observations generated by the coherent model in Section~\ref{sec:physical_model}, our goal is to estimate the latent reflectivity $x$ without clean ground truth. Fig.~2 summarizes the proposed self-supervised framework. The reconstruction network is defined as
\begin{equation}
\hat{x}=f_{\theta}(y).
\end{equation}
This inverse problem is ill-conditioned because coherent speckle introduces signal-dependent stochastic fluctuations, while the forward operator induced by the sonar PSF suppresses high-frequency reflectivity structure. Therefore, training must constrain the reconstruction in the measurement domain while regularizing the latent reflectivity estimate.

\subsection{Paired Observation Construction}
\label{sec:registration_assumption}

\zx{Multi-frame sonar sequences provide the opportunity to exploit repeated observations when viewing an approximately static scene. To this end, we select and register pairs of frames to form observations $(y_1,y_2^{\mathrm{reg}})$ that share the same latent reflectivity but have different coherent speckle realizations. The paired frame can be selected from a short temporal window around $y_1$. For short acquisition intervals, the dominant inter-frame motion is approximated by a two-dimensional translation $\Delta\boldsymbol{\rho}=(\Delta\rho_1,\Delta\rho_2)$, estimated by phase correlation \cite{registration,registration_yvan}. We denote the resulting translation warp by $W_{2\to1}$, so that $y_2^{\mathrm{reg}}=W_{2\to1}(y_2^{\mathrm{raw}})$ aligns the paired frame to $y_1$. After registration, the paired frames are assumed to have aligned conditional means,}
\begin{equation}
\mathbb{E}\!\left[
y_2^{\mathrm{reg}}[\boldsymbol{\rho}]
\,\middle|\, x
\right]
\approx
\mathbb{E}\!\left[
y_1[\boldsymbol{\rho}]
\,\middle|\, x
\right],
\end{equation}
\zx{where $\boldsymbol{\rho}=(\rho_1,\rho_2)$ denotes a discrete range-angle image coordinate. For notational simplicity, we write $y_2$ for $y_2^{\mathrm{reg}}$ below.}
\zx{A representative before-and-after registration check is provided in the \supp{Supplementary Section~S1} to illustrate the structural alignment obtained by this procedure.}

\subsection{Statistical Assumptions and Stochastic Forward Model}
\label{sec_bridge}
For a registered pair observing the same local scene, we model
\begin{equation}
y_k=\left|(\sqrt{x}\cdot g_k)*h\right|^2,
\qquad k\in\{1,2\},
\end{equation}
where $x$ is the shared latent reflectivity, $h$ is the coherent PSF, and $g_k$ is the coherent speckle modulation field of the $k$-th observation.
We assume approximate scene stationarity after registration and reduced inter-frame speckle correlation. For the paired-loss analysis, this is expressed through the conditional-mean relation
\begin{equation}
\label{eq:assumption}
\mathbb{E}[y_2\mid x,y_1]\approx \mathbb{E}[y_2\mid x]=\mu(x),
\qquad
\mathbb{E}[y_1\mid x]=\mu(x).
\end{equation}
Exact conditional independence, $y_1\perp y_2\mid x$, is a sufficient but stronger condition for this relation.

For a fixed reflectivity $x$, the measurement-domain target under the mean squared error criterion is
\begin{equation}
\mu(x)\triangleq \mathbb{E}_{g}[y\mid x].
\end{equation}
Let \(v\) denote the coherent field after PSF convolution:
\begin{equation}
v=(\sqrt{x}\cdot g)*h,
\qquad
y=|v|^2.
\end{equation}
\zx{For a two-dimensional index $p=(p_1,p_2)\in\Omega$ and $q=(q_1,q_2)\in\Omega$, the spatial convolution can be written as}
\begin{equation}
v_p=\sum_{q\in\Omega} h_{p-q}\sqrt{x_q}g_q .
\end{equation}
Then,
\begin{align}
\mu_p(x)
&=
\mathbb{E}_{g}[|v_p|^2\mid x]  \\
&=
\sum_{q\in\Omega}\sum_{r\in\Omega}
h_{p-q}h_{p-r}^{*}
\sqrt{x_qx_r}\,
\mathbb{E}[g_qg_r^{*}].
\end{align}
Using the spatial whiteness and unit second moment of the coherent speckle modulation field \cite{array_signal_processing},
\begin{equation}
\mathbb{E}_{g}[g_qg_r^{*}]=\delta_{qr}.
\end{equation}
where $\delta_{qr}$ is the Kronecker delta. Hence, all cross terms vanish, giving
\begin{equation}
\mu_p(x)=\sum_q x_q |h_{p-q}|^2.
\end{equation}
\zx{Thus, under the local shift-invariant approximation, despite the stochastic observation model being nonlinear in $x$ and $g$, the conditional mean can be expressed as a linear function of $x$:}
\begin{equation}
\label{eq:mu_result}
\mu(x)=A x=x*|h|^2,
\end{equation}
\zx{where $A$ is the conditional-mean forward operator induced by the coherent observation model.}

\zx{The same forward model is also used to simulate sonar measurements from a candidate reflectivity image $z$. We define the stochastic coherent simulator}
\begin{equation}
S_g(z):=\left|(\sqrt{z}\cdot g)*h\right|^2.
\label{eq:stochastic_forward_simulator}
\end{equation}
\zx{Following the speckle-averaging principle in coherent imaging \cite{goodman_speckle_book}, we use a $J$-look averaged simulator}
\begin{equation}
\bar{S}_J(z):=\frac{1}{J}\sum_{j=1}^{J} S_{g_j}(z),
\label{eq:avg_simulator_ei}
\end{equation}
where $g_1,\dots,g_J$ are independent speckle fields. Provided each simulated look satisfies $\mathbb{E}_{g_j}[S_{g_j}(z)\mid z]=A z$, the averaged simulator preserves the same conditional mean,
\begin{equation}
\mathbb{E}_{g_1,\ldots,g_J}\!\left[\bar{S}_J(z)\mid z\right] = A z,
\label{eq:avg_sim_mean}
\end{equation}

Equivalently, writing $\varepsilon_j(z)=S_{g_j}(z)-Az$, we have $\mathbb{E}[\varepsilon_j(z)\mid z]=0$ and $\bar{S}_J(z)-Az=J^{-1}\sum_{j=1}^{J}\varepsilon_j(z)$. 
Independence of the simulated looks eliminates the cross terms in expectation, giving
\begin{equation}
\begin{aligned}
&\mathbb{E}_{g_1,\ldots,g_J}\!\left[
\left\|
\bar{S}_J(z)-A z
\right\|_2^2
\,\middle|\, z
\right] \\
&\qquad =
\frac{1}{J}
\mathbb{E}_{g}\!\left[
\left\|
S_g(z)-A z
\right\|_2^2
\,\middle|\, z
\right].
\end{aligned}
\label{eq:avg_sim_var}
\end{equation}
Thus, $J$ controls the strength of the simulated speckle while preserving the conditional mean learning goal. This supports the interpretation of $\bar{S}_J$ as a $J$-look simulated sonar measurement: $J=1$ corresponds to a single-look simulated observation, whereas $\bar{S}_J(z)$ converges in mean square to $A(z)$ as $J\to\infty$ under the independent-look simulator.

\subsection{Physics Consistency Learning}
\label{sec:PCLoss}

\zx{Our objective is to learn a reconstruction network that approximates the conditional mean of the latent reflectivity,
\begin{equation}
\hat{x}=f_{\theta}(y_1)\approx \mathbb{E}[x\mid y_1].
\end{equation}}
However, clean samples of $x$ are not available. Using the registered-pair assumptions in Section~\ref{sec_bridge} and the conditional mean relation in \eqref{eq:mu_result}, we train through the measurement-domain prediction $A f_{\theta}(y_1)$. The physics consistency loss is
\begin{equation}
\label{eq:pc_loss}
\mathcal{L}_{\mathrm{PC}}(\theta)
=
\mathbb{E}_{(y_1,y_2)\sim p_{\mathrm{pair}}}
\left[
\left\|
y_2-A f_{\theta}(y_1)
\right\|_2^2
\right],
\end{equation}
where $p_{\mathrm{pair}}$ denotes the distribution of registered paired observations. Under the assumptions above, this loss decomposes as
\begin{align}
\label{eq:pc_main_result}
\mathcal{L}_{\mathrm{PC}}(\theta)
&=
\mathbb{E}_{x,y_1}
\left[
\left\|
A x-A f_{\theta}(y_1)
\right\|_2^2
\right]
+C ,
\end{align}
where
\begin{equation}
\label{eq:pc_constant}
C=
\mathbb{E}_{x}
\left[
\mathbb{E}_{y_2\mid x}
\left[
\|y_2-\mu(x)\|_2^2
\right]
\right]
\end{equation}
is independent of $\theta$.

Under the idealized condition $\mathbb{E}[y_2\mid x,y_1]=Ax$, the cross term in the squared-loss expansion vanishes, yielding \eqref{eq:pc_main_result}, with the network-independent constant defined in \eqref{eq:pc_constant}.
Thus, minimizing $\mathcal{L}_{\mathrm{PC}}$ matches the conditional mean projected onto the measurement domain, rather than fitting a single noisy realization.

\zx{The minimizers are not necessarily unique in the reflectivity domain. If $A^\dagger$ denotes the Moore--Penrose pseudoinverse, a solution can be written as}
\begin{equation}
f_{\theta}(y_1)
=
A^\dagger A\,\mathbb{E}[x\mid y_1]
+
\nu(y_1),
\qquad
\nu(y_1)\in\mathrm{null}(A),
\end{equation}
\zx{consistent with the equivariant splitting ambiguity of inverse problems \cite[Lemma~1]{EIsplitting}. In practice, even when $A$ is full rank, it is a blur operator and can be severely ill-conditioned. Therefore, the physics consistency loss alone weakly determines high-frequency reflectivity components, motivating the equivariant regularization introduced next.}

\subsection{Robust Equivariant Imaging with Stochastic Forward Simulation}
\label{sec_ei}
\zx{To further constrain the latent reflectivity, we adapt robust equivariant imaging (REI) \cite{robustEI} to the coherent speckle noise model introduced above. REI enforces consistency of the reconstruction system under transformations that preserve plausible reflectivity structure. Let $T_\tau$ denote such a transformation, for example a small rotation or shift. Given}
\begin{equation}
\hat{x}=f_\theta(y_1),
\qquad
z=T_\tau \hat{x},
\end{equation}
\zx{the REI branch requires the network to recover $z$ from a simulated sonar measurement generated using the $J$-look coherent simulator $\bar{S}_J$. The sonar-adapted REI loss is}

\begin{equation}
\label{eq:rei_loss_final}
\begin{aligned}
\mathcal{L}_{\mathrm{REI}}^{(J)}(\theta)
=
\mathbb{E}_{\substack{
y_1,\tau,
g_1,\dots,g_J
}}
\Big[
\big\|
&f_\theta\!\left(\bar{S}_J(T_\tau f_\theta(y_1))\right) \\
&- T_\tau f_\theta(y_1)
\big\|_2^2
\Big].
\end{aligned}
\end{equation}

The final training objective is
\begin{equation}
\mathcal{L}_{\mathrm{total}}(\theta)
=
\mathcal{L}_{\mathrm{PC}}(\theta)
+
\alpha\,\mathcal{L}_{\mathrm{REI}}^{(J)}(\theta),
\label{eq:total_loss_final}
\end{equation}
where $\alpha>0$ controls the REI regularization strength.

\section{Experiments}

We evaluate the proposed framework on two real multibeam imaging sonar datasets acquired with a Tritech Gemini 720ik sonar.
\zx{In the absence of clean reflectivity ground truth, real-sonar evaluation cannot directly measure latent reflectivity accuracy. However, when registered pairs of corrupted sonar observations are available, the same Noise2Noise principle used for training provides a paired-frame proxy for measurement-domain predictive consistency. We therefore assess restoration using complementary evidence: the proposed paired forward prediction error (PFPE) measures measurement-domain forward-prediction consistency, while equivalent number of looks (ENL), edge preservation index (EPI), and visual comparisons assess speckle suppression, edge preservation, and structural plausibility. In this setting, deblurring is evaluated indirectly through consistency with the sonar forward model and preservation of visible structures.}

\subsection{Evaluation Metrics}
\label{sec:evaluation_metrics}
The PFPE metric tests whether a restoration from one frame can predict a registered paired observation.
\zx{Let $y_1$ be the input frame, $y_2$ its registered paired observation, and $\hat{x}$ the restored output of an evaluated method. For the proposed method, $\hat{x}=f_\theta(y_1)$. We define}
\begin{equation}
\mathrm{PFPE}(y_1,y_2)
=
\left\|A \hat{x}-y_2\right\|_2^2.
\label{eq:pfpe}
\end{equation}
The aggregate PFPE improvement is computed from the ratio of empirical
mean errors as
\begin{equation}
\mathrm{PFPE}_{\mathrm{imp}}^{\mathrm{agg}}
=100\left(1-\frac{\overline{\mathrm{PFPE}(y_1,y_2)}}{\overline{\|y_1-y_2\|_2^2}
}
\right),
\label{eq:pfpe_imp_agg}
\end{equation}
Here, squared norms are normalized by the number of evaluated
image pixels, and the overline denotes the empirical mean
over all evaluated test pairs. Tables report PFPE as mean $\pm$
standard deviation across these pairs.

\zx{Lower PFPE and higher $\mathrm{PFPE}_{\mathrm{imp}}^{\mathrm{agg}}$ scores indicate better agreement under the fixed paired-frame protocol. Under the idealized conditional-mean condition in Section~\ref{sec_bridge}, expected PFPE equals expected conditional-mean measurement error plus a reconstruction-independent constant. 
Unlike single-frame consistency $\|A f_\theta(y_1)-y_1\|_2^2$, the held-out criterion is less biased toward preserving input-specific speckle because the prediction is compared with a registered neighbouring observation. 
All restored outputs are scored using the same sonar operator $A$ to predict a registered paired observation, so PFPE is interpreted under a common latent-reflectivity hypothesis rather than as a full-reference quality metric or an evaluation of each baseline's native prediction target. We therefore interpret PFPE jointly with ENL, EPI, $D_{\mathrm{EPI}}$, and visual results.}

\zx{We also report ENL and edge preservation index deviation ($D_{\mathrm{EPI}}$) as complementary no-reference indicators. These metrics characterize homogeneous-region smoothing and edge-strength preservation, respectively, but they do not measure reconstruction accuracy in the absence of ground truth.} ENL measures speckle suppression in homogeneous regions,
\begin{equation}
    \mathrm{ENL}
    =
    \frac{\mu_{\hat{x},\Omega_{\mathrm{hom}}}^2}
    {\sigma_{\hat{x},\Omega_{\mathrm{hom}}}^2},
\end{equation}
where $\Omega_{\mathrm{hom}}$ denotes homogeneous regions and $\mu_{\hat{x},\Omega_{\mathrm{hom}}}$ and $\sigma_{\hat{x},\Omega_{\mathrm{hom}}}^2$ are the mean and variance of $\hat{x}$ within these regions. Higher ENL indicates stronger speckle suppression, but may favor oversmoothing.

EPI measures relative edge strength in selected edge regions:
\begin{equation}
\mathrm{EPI}_n
=
\frac{
\sum_{(i,j)\in\Omega_{n,\mathrm{edge}}}
\left|\nabla \hat{x}_n(i,j)\right|
}{
\sum_{(i,j)\in\Omega_{n,\mathrm{edge}}}
\left|\nabla y_n(i,j)\right|
},
\label{eq:epi}
\end{equation}
where $\Omega_{n,\mathrm{edge}}$ is selected from the observed image $y_n$
and held fixed across methods, and $\nabla$ denotes a finite-difference
gradient operator. We report the mean per-image deviation from unity:
\begin{equation}
D_{\mathrm{EPI}}
=
\frac{100}{N}
\sum_{n=1}^{N}
\left|1 - \mathrm{EPI}_n\right|,
\label{eq:epi_deviation}
\end{equation}
where $N$ is the number of evaluated images. A smaller
$D_{\mathrm{EPI}}$ indicates closer agreement of edge strength with
the observed images.

For ENL, homogeneous ROI candidates are required to have sufficient mean intensity, valid-pixel support, and low Sobel gradient energy, and the retained candidates are ranked by coefficient of variation; for EPI and $D_{\mathrm{EPI}}$, edge ROIs are selected separately as the patches with the highest Sobel gradient energy. All ROIs are selected from the input observation and then fixed for all compared methods. This avoids selecting regions according to the restored output and reduces manual selection bias.

\subsection{Datasets}
\label{sec:datasets}

We use two datasets acquired with a Tritech Gemini 720ik sonar. \zx{The sensor outputs intensity images formed from 512 beams, operates at a centre frequency of 720~kHz, and provides a $120^\circ$ field of view.} Raw measurements are represented as range-beam images in polar coordinates and treated as single-channel intensity images.

\textbf{Boat.}
The Boat dataset contains complex underwater scenes with cluttered object responses and strong background speckle, seabed texture, acoustic shadows, and heterogeneous scattering from object boundaries and surrounding structures. These factors make the dataset challenging for joint deblurring and despeckling, since the method must suppress coherent speckle while preserving structural details and boundaries. It was acquired with a maximum imaging range of $50~\mathrm{m}$ and a frame rate of $3$ frames per second. The dataset contains $1455$ raw frames.

\textbf{Underwater Structure.}
The Underwater Structure dataset contains more regular structural scenes and provides a complementary evaluation under relatively stable underwater inspection conditions. It was acquired with a maximum imaging range of $25.5~\mathrm{m}$ and a frame rate of $10$ frames per second. The dataset contains $1282$ raw frames.

\subsection{Implementation Details}
\label{sec:implementation}
For each raw sequence, we use temporally contiguous training, validation, and test blocks with an approximate $8:1:1$ ratio before fixed $512\times512$ patch extraction. A small buffer of unused frames is inserted between adjacent blocks, and paired-frame samples are formed only within the same split. This prevents temporally adjacent or registered pairs from crossing the train/validation/test boundaries.
During testing, the fully convolutional network is applied to individual sonar frames. We compare the proposed method with representative model-based despeckling methods, including PPB \cite{PPB} and SAR-BM3D \cite{SAR_BM3D}, and recent learning-based self-supervised baselines, including Noise2Void \cite{noise2void}, Blind2Unblind (B2UB) \cite{Blind2unblind}, LG-BPN \cite{LG_BPN}, SS-BSN \cite{SS_BSN}, SEGSID \cite{SEGSID}, and PI-SSD \cite{RegN2N}.

\zx{All paired-frame experiments use a fixed main protocol with frame gap 1. The gap is chosen as the smallest reliable gap that reduces inter-frame correlation while preserving scene match and registration quality. Pairs are retained only when the estimated translation satisfies $|\Delta \rho_1|\leq 10$ pixels and $|\Delta \rho_2|\leq 5$ pixels, which removes pairs with large apparent motion, scene mismatch, or failed registration. This yields 1328 and 1137 retained registered pairs for the Boat and Underwater Structure datasets, respectively, under the final pairing protocol. The full gap-correlation statistics and displacement diagnostics are reported in \supp{Supplementary Section~S2}. All methods are evaluated using the same fixed test pairs.}

The PSF is generated from the Gemini 720ik geometry and parameters reported in Section~\ref{sec:datasets}, together with a sound speed of $1540~\mathrm{m/s}$ and a 96-element receive aperture with half-wavelength spacing.

For the network, we use a standard U-Net backbone implemented with \texttt{deepinv} \cite{Deepinv}. This compact four-scale U-Net keeps the inference model lightweight and is chosen to support future low-latency deployment on robotic sonar platforms. All models are trained with Adam using a learning rate of $1\times10^{-4}$, a weight decay of $5\times10^{-7}$, and a batch size of $2$. The checkpoint with the lowest validation loss is selected for testing. Unless otherwise stated, the main experiments use $J=6$, with $T_\tau$ implemented as random rotations with a maximum angle of $20^\circ$ for REI regularization. The REI weight is set to $\alpha=0.03$ for Boat and $\alpha=0.05$ for Underwater Structure, based on validation performance. This dataset-specific choice reflects the different registration accuracy and structural complexity of the two datasets, and no test-set results were used for this selection. Experiments are implemented in PyTorch~2.2.2 and run on a single NVIDIA RTX~3090 GPU.

\subsection{Performance on Boat Dataset}
\label{sec:boat_performance}

\begin{figure}[t]
    \raggedright
     \includegraphics[width=0.98\linewidth]{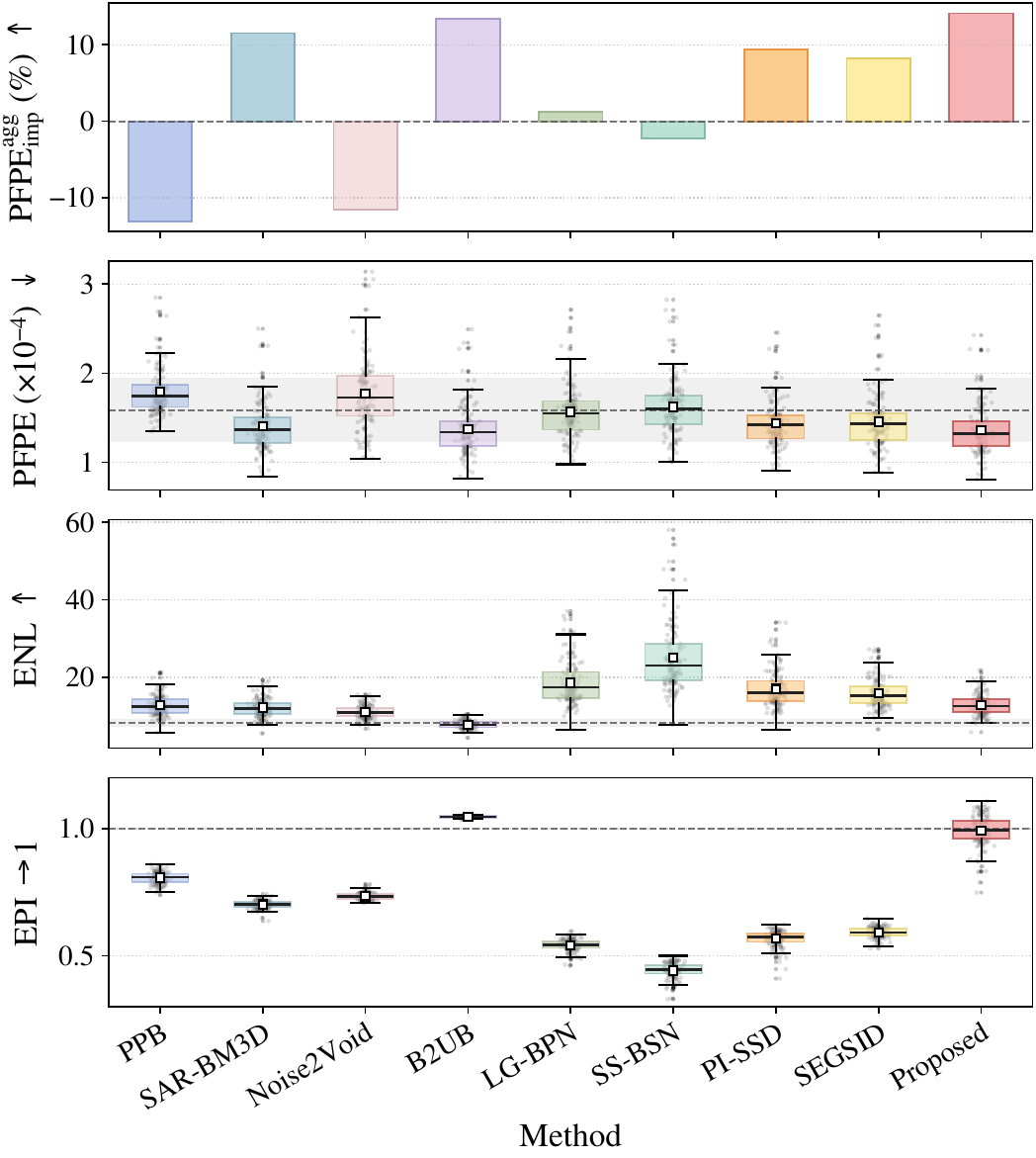}
\caption{
The top panel shows $\mathrm{PFPE}_{\mathrm{imp}}^{\mathrm{agg}}$
relative to the noisy baseline; the remaining three panels show
distributions of per-pair PFPE and per-image ENL and EPI
on the Boat dataset.
In the three boxplots, boxes show interquartile ranges, horizontal lines indicate medians, and dots denote individual test images.
Dashed gray lines and shaded gray bands indicate the corresponding noisy means and mean $\pm$ one standard deviation, respectively. 
}
    \label{fig:boat_metric_boxplots}
\end{figure}

\noindent\textbf{Quantitative evaluation.}
The Boat results are reported in the first four metric columns of Table~\ref{tab:main_results}, and the corresponding metric distributions are shown in Fig.~\ref{fig:boat_metric_boxplots}. Additional qualitative examples and complete metric values are provided in \supp{Supplementary Section~S3}, including \supp{Fig.~7} and \supp{Table~V}. PFPE measures the raw held-out forward-prediction error, while $\mathrm{PFPE}_{\mathrm{imp}}^{\mathrm{agg}}$ reports the aggregate improvement over the noisy registered-pair baseline.

\zx{The proposed method achieves the lowest PFPE and the highest aggregate PFPE improvement of $14.10\%$ under the stated common latent-reflectivity scoring protocol, indicating the strongest overall forward-prediction consistency on the Boat dataset.} Blind2Unblind (B2UB) obtains a similar aggregate PFPE improvement of $13.39\%$, but its ENL decreases from the raw image value of $8.18$ to $7.74$, indicating limited speckle suppression. \zx{In contrast, the proposed method increases ENL to $12.81$ while keeping $D_{\mathrm{EPI}}$ among the lowest restored results.} Methods with larger ENL, such as SS-BSN, LG-BPN, and PI-SSD, show much larger edge-strength deviation and lower PFPE improvement, suggesting that their higher smoothness is associated with oversmoothing rather than stronger forward-prediction agreement.
\zx{The ablation results further show why PFPE should not be interpreted in isolation. As reported in \supp{Supplementary Section~S5-A} and \supp{Supplementary Table~VII}, a physics-consistency-only ($\mathcal{L}_{\mathrm{PC}}$-only) variant improves PFPE but yields nearly zero ENL and extremely large $D_{\mathrm{EPI}}$, demonstrating that forward consistency alone does not ensure a plausible reconstruction. This supports the use of REI regularization and motivates the multi-metric evaluation protocol: PFPE is treated as a measurement-domain consistency metric and interpreted together with ENL, $D_{\mathrm{EPI}}$, and visual structural evidence.}

\begin{figure*}[t]
    \centering
    \captionsetup[subfigure]{font=footnotesize}
    \newcommand{\boatfigwB}{0.18\textwidth}

    \begin{subfigure}[t]{\boatfigwB}
        \centering
        \includegraphics[width=\linewidth]{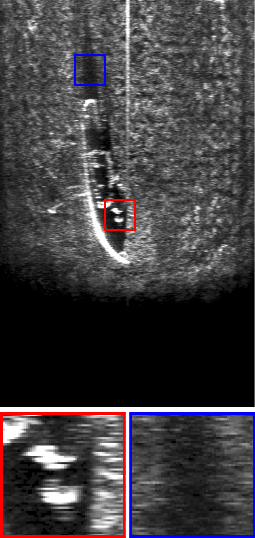}
        \caption{Noisy}
    \end{subfigure}
    \hfill
    \begin{subfigure}[t]{\boatfigwB}
        \centering
        \includegraphics[width=\linewidth]{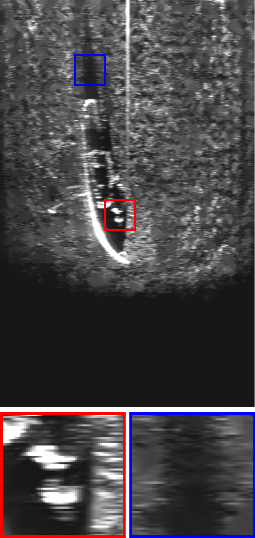}
        \caption{PPB}
    \end{subfigure}
    \hfill
    \begin{subfigure}[t]{\boatfigwB}
        \centering
        \includegraphics[width=\linewidth]{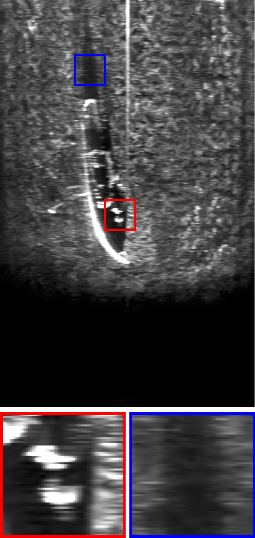}
        \caption{SAR-BM3D}
    \end{subfigure}
    \hfill
    \begin{subfigure}[t]{\boatfigwB}
        \centering
        \includegraphics[width=\linewidth]{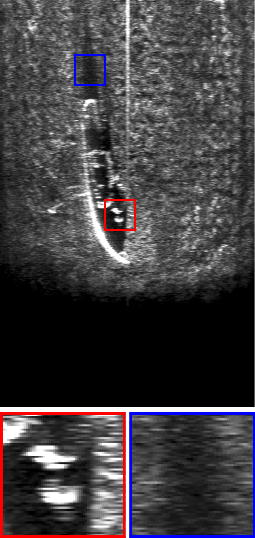}
        \caption{Blind2Unblind}
    \end{subfigure}
    \hfill
    \begin{subfigure}[t]{\boatfigwB}
        \centering
        \includegraphics[width=\linewidth]{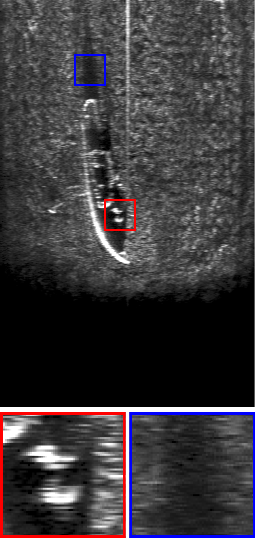}
        \caption{Noise2Void}
    \end{subfigure}

    \vspace{0.6em}

    \begin{subfigure}[t]{\boatfigwB}
        \centering
        \includegraphics[width=\linewidth]{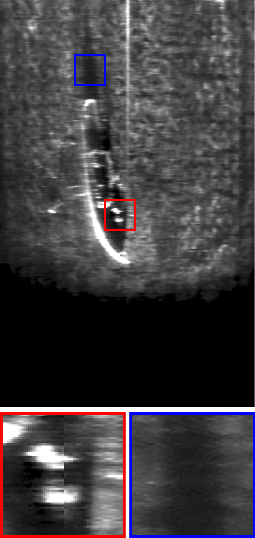}
        \caption{LG-BPN}
    \end{subfigure}
    \hfill
    \begin{subfigure}[t]{\boatfigwB}
        \centering
        \includegraphics[width=\linewidth]{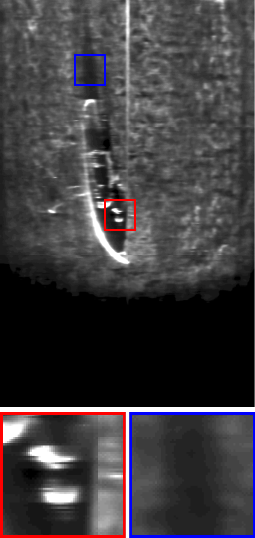}
        \caption{SS-BSN}
    \end{subfigure}
    \hfill
    \begin{subfigure}[t]{\boatfigwB}
        \centering
        \includegraphics[width=\linewidth]{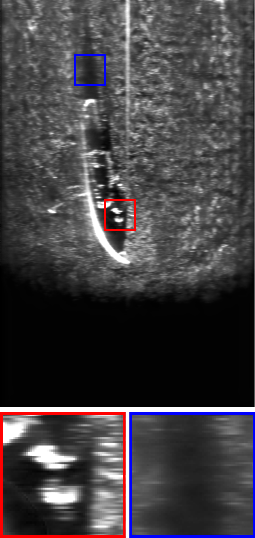}
        \caption{PI-SSD}
    \end{subfigure}
    \hfill
    \begin{subfigure}[t]{\boatfigwB}
        \centering
        \includegraphics[width=\linewidth]{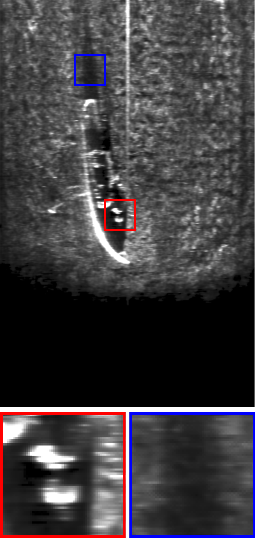}
        \caption{SEGSID}
    \end{subfigure}
    \hfill
    \begin{subfigure}[t]{\boatfigwB}
        \centering
        \includegraphics[width=\linewidth]{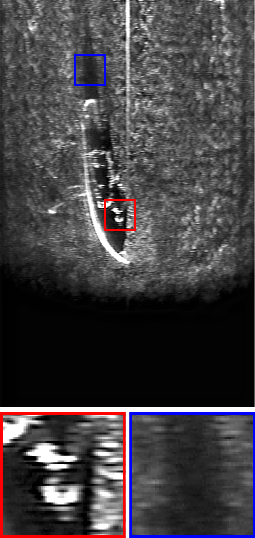}

        \caption{Proposed}
    \end{subfigure}

        \caption{
Qualitative comparison on the Boat dataset. Red and blue boxes show zoomed regions around high-reflectivity structures and background-shadow transitions. The proposed method reduces speckle while preserving separable local structures and clearer object-background transitions compared with representative model-based and self-supervised methods.
}
    \label{fig:boat_visual_results_2}
\end{figure*}

\noindent\textbf{Qualitative evaluation.}
Fig.~\ref{fig:boat_visual_results_2} shows representative visual comparisons on Boat. PPB and SAR-BM3D reduce speckle but blur local structures, while several learning-based baselines either retain granular texture or over-smooth object boundaries. The proposed method suppresses background speckle while preserving clearer separation of high-reflectivity structures in the zoomed regions. These observations agree with the quantitative results, where the proposed method achieves the best aggregate PFPE improvement while maintaining low edge-strength deviation.

\begin{table*}[t]
\centering
\caption{
Quantitative comparison on the Boat and Underwater Structure datasets. PFPE is reported in units of $10^{-4}$, and $D_{\mathrm{EPI}}$ is reported in percent. $\mathrm{PFPE}_{\mathrm{imp}}^{\mathrm{agg}}$ denotes the aggregate percentage improvement over the noisy-frame baseline. The lowest PFPE and highest $\mathrm{PFPE}_{\mathrm{imp}}^{\mathrm{agg}}$ are highlighted in bold.
ENL and $D_{\mathrm{EPI}}$ are complementary indicators of speckle suppression and edge-strength deviation, respectively. 
}
\label{tab:main_results}
\begin{tabular*}{\textwidth}{@{\extracolsep{\fill}} l c c c c c c c c}
\toprule
&
\multicolumn{4}{c}{Boat}
& \multicolumn{4}{c}{Underwater Structure} \\
\cmidrule(lr){2-5}
\cmidrule(lr){6-9}
Method
& $\mathrm{PFPE}$ $\downarrow$ 
& $\mathrm{PFPE}_{\mathrm{imp}}^{\mathrm{agg}}$ $\uparrow$ 
& ENL $\uparrow$
& $D_{\mathrm{EPI}}$ $\downarrow$
& $\mathrm{PFPE}$ $\downarrow$ 
& $\mathrm{PFPE}_{\mathrm{imp}}^{\mathrm{agg}}$ $\uparrow$ 
& ENL $\uparrow$
& $D_{\mathrm{EPI}}$ $\downarrow$ \\
\midrule
Noisy
& $ 1.59 \pm 0.36$
& $0.00$
& $8.18 \pm 1.12$
& $0.00 \pm 0.00$
& $6.72 \pm 3.06$
& $0.00$
& $3.38 \pm 0.36$
& $0.00 \pm 0.00$ \\

\midrule
PPB \cite{PPB}
& $ 1.79 \pm 0.26$
& $-13.06$
& $12.73 \pm 2.71$
& $19.32 \pm 2.28$
& $ 7.33 \pm 2.34$
& $-9.01$
& $6.98 \pm 1.36$
& $27. 17\pm 4.16$ \\

SAR-BM3D \cite{SAR_BM3D}
& $1.40 \pm 0.29$
& $11.53$
& $12.14 \pm 2.40$
& $29.87 \pm 1.60$
& $6.72 \pm 2.18$
& $-0.02$
& $4.71 \pm 0.59$
& $25.42 \pm 1.05$ \\

\midrule
Noise2Void \cite{noise2void}
& $1.77 \pm 0.40$
& $-11.59 $
& $11.03 \pm 1.62$
& $26.61 \pm 1.36$
& $6.82 \pm 2.24$
& $-1.49 $
& $4.09 \pm 0.46$
& $17.20 \pm 0.56$ \\

B2UB \cite{Blind2unblind}
& $1.37 \pm 0.29$
& $13.39$
& $7.74 \pm 1.03$
& $4.68 \pm 0.36$
& $6.57\pm 2.18$
& $2.25$
& $3.36 \pm 0.35$
& $1.11 \pm 0.13$ \\

LG-BPN \cite{LG_BPN}
& $1.57 \pm 0.31$
& $1.19 $
& $18.57 \pm 5.60$
& $45.89 \pm 2.27$
& $7.80 \pm 2.35$
& $-16.05 $
& $7.11 \pm 1.35$
& $14.19 \pm 2.16$ \\

SS-BSN \cite{SS_BSN}
& $1.62 \pm 0.32$
& $-2.24$
& $25.04 \pm 8.79$
& $55.92 \pm 3.06$
& $8.85 \pm 2.84$
& $-31.66$
& $15.50 \pm 5.22$
& $71.24 \pm 3.88$ \\

PI-SSD \cite{RegN2N}
& $1.44 \pm 0.27$
& $9.39$
& $16.97 \pm 4.81$
& $43.22 \pm 3.24$
& $7.21 \pm 2.25$
& $-7.36$
& $7.76 \pm 2.02$
& $51.29 \pm 2.19$ \\

SEGSID \cite{SEGSID}
& $1.46 \pm 0.31 $
& $8.20$
& $15.83 \pm 3.81$
& $40.82 \pm  2.16$
& $ 6.86\pm 2.22$
& $-2.02$
& $4.74 \pm 0.66$
& $28.86 \pm 1.44$ \\

\midrule

Proposed 
& \best{$1.36 \pm 0.28$}
& \best{$14.10$}
& $12.81 \pm 2.68$
& $4.65 \pm 4.13$
& \best{$ 6.43\pm 2.08$}
& \best{$4.36$}
& $6.37 \pm 2.84$
& $6.83 \pm 2.95$ \\

\bottomrule
\end{tabular*}
\end{table*}

\subsection{Performance on Underwater Structure Dataset}
\label{sec:underwater_structure_performance}

\noindent\textbf{Quantitative evaluation.}
The Underwater Structure results are reported in the last four metric columns of Table~\ref{tab:main_results}. Additional qualitative examples and complete metric values for this dataset are provided in \supp{Supplementary Section~S4}, including \supp{Fig.~8} and \supp{Table~VI}.

The proposed method achieves the lowest PFPE and the highest aggregate PFPE improvement of $4.36\%$ under the latent-reflectivity hypothesis, indicating the strongest overall agreement with the paired forward-prediction criterion on this dataset. 
As in the Boat experiment, methods with substantially higher ENL, such as SS-BSN, PI-SSD, and LG-BPN, also show much larger $D_{\mathrm{EPI}}$ values, indicating that stronger smoothing does not necessarily preserve observed edge strength. 
In contrast, the proposed method provides the best aggregate PFPE improvement while keeping $D_{\mathrm{EPI}}$ relatively low, showing a favourable balance between forward-prediction consistency, speckle suppression, and structure preservation.

\begin{figure*}[t]
    \centering
    \captionsetup[subfigure]{font=footnotesize}

    \begin{subfigure}[t]{0.19\textwidth}
        \centering
        \includegraphics[width=\linewidth]{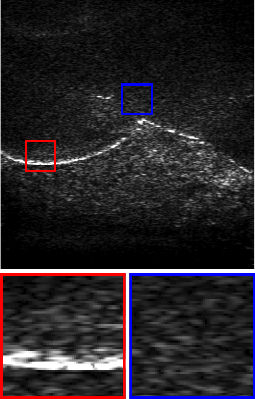}
        \caption{Noisy}
    \end{subfigure}
    \hfill
    \begin{subfigure}[t]{0.19\textwidth}
        \centering
        \includegraphics[width=\linewidth]{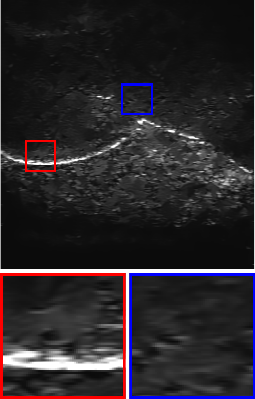}
        \caption{PPB}
    \end{subfigure}
    \hfill
    \begin{subfigure}[t]{0.19\textwidth}
        \centering
        \includegraphics[width=\linewidth]{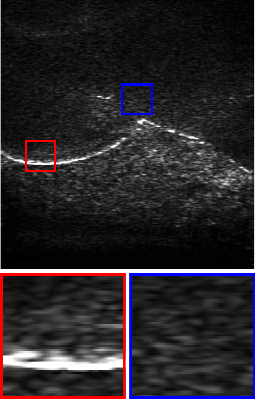}
        \caption{SAR-BM3D}
    \end{subfigure}
     \hfill
    \begin{subfigure}[t]{0.19\textwidth}
        \centering
        \includegraphics[width=\linewidth]{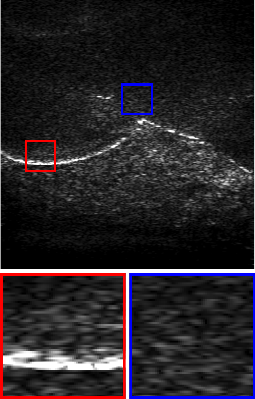}
        \caption{Blind2Unblind}
    \end{subfigure}
     \hfill
    \begin{subfigure}[t]{0.19\textwidth}
        \centering
        \includegraphics[width=\linewidth]{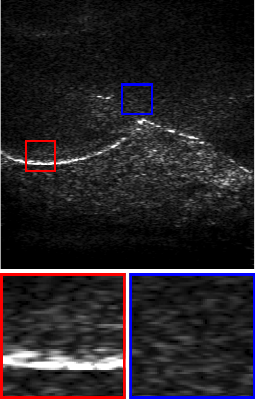}
        \caption{Noise2Void}
    \end{subfigure}
    \hfill

       \vspace{0.5em} 


    \begin{subfigure}[t]{0.19\textwidth}
        \centering
        \includegraphics[width=\linewidth]{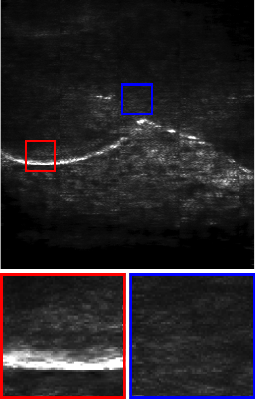}
        \caption{LG-BPN}
    \end{subfigure}
    \hfill
    \begin{subfigure}[t]{0.19\textwidth}
        \centering
        \includegraphics[width=\linewidth]{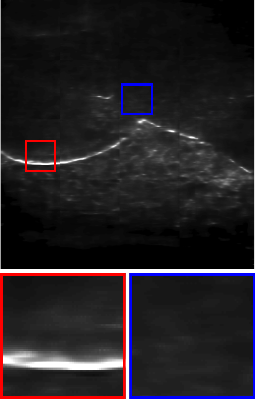}
        \caption{SS-BSN}
    \end{subfigure}
    \hfill
    \begin{subfigure}[t]{0.19\textwidth}
        \centering
        \includegraphics[width=\linewidth]{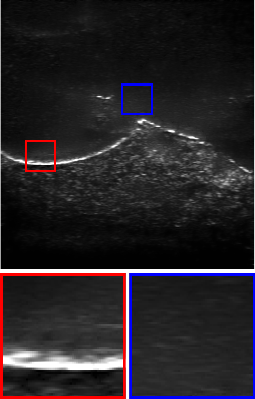}
        \caption{PI-SSD}
    \end{subfigure}
    \hfill
    \begin{subfigure}[t]{0.19\textwidth}
        \centering
        \includegraphics[width=\linewidth]{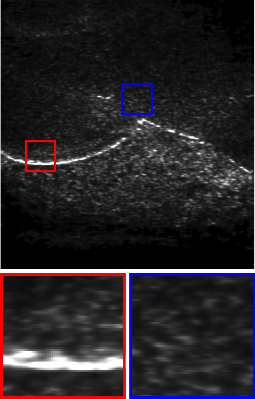}
        \caption{SEGSID}
    \end{subfigure}
    \hfill
    \begin{subfigure}[t]{0.19\textwidth}
        \centering
        \includegraphics[width=\linewidth]{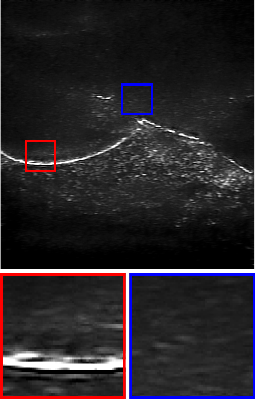}
        \caption{Proposed}
    \end{subfigure}
    \hfill
    \caption{
Qualitative comparison on the Underwater Structure dataset. Red and blue boxes show zoomed regions around structural boundaries and background regions. The proposed method suppresses background speckle while preserving clearer structural continuity and weak local responses compared with representative baselines.
}
\label{fig:underwater_structure_visual_results}

\end{figure*}

\noindent\textbf{Qualitative evaluation.}
Fig.~\ref{fig:underwater_structure_visual_results} shows representative visual comparisons on the Underwater Structure dataset. 
Model-based methods retain residual speckle or blur weak structural details, while several self-supervised baselines produce smoother backgrounds at the cost of less distinct object boundaries. 
The proposed method suppresses background speckle while preserving clearer structural continuity, especially in the zoomed regions where the boundary between the underwater structure and the surrounding background remains more distinguishable.

\subsection{Computational Cost}
\label{sec:computational_cost}

\zx{Table~\ref{tab:computational_cost} reports the computational cost on $512 \times 512$ images. Parameters and MACs are computed at the model level for learning-based methods, while time per image is measured using the actual test-time inference path used to generate the reported results. For the proposed method, only the trained reconstruction network is used at inference; the paired observation branch, stochastic simulator, and REI loss are used only during training.}

\zx{The proposed method has the same backbone as PI-SSD and therefore the same parameter count and MACs, with a measured inference time of $0.043$ s per image. This corresponds to approximately $23$ frames per second, which is faster than the acquisition rates of both datasets considered here: $3$ frames per second for Boat and $10$ frames per second for Underwater Structure. This suggests that the physics-consistent self-supervised training objective improves restoration quality without requiring a heavier test-time model, and remains compatible with future real-time multibeam sonar processing.}

\begin{table}[t]
\centering
\caption{
Computational cost on Boat images. MACs are reported using the model-level protocol, and time per image is measured using the actual inference path.}
\label{tab:computational_cost}
\begin{tabular*}{\columnwidth}{@{\extracolsep{\fill}}lccc}
\toprule
Method & Params (M) & MACs (G) & Time/img (s) $\downarrow$ \\
\midrule
PPB & -- & -- & $151.219$\\
SAR-BM3D & -- & -- & $324.225$ \\
Noise2Void & $0.80$ & $52.26$ & $0.147$ \\
B2UB & $1.10$ & $73.94$ & $0.018$ \\
LG-BPN & $6.16$ & $1605.23$ & $13.789$ \\
SS-BSN & $6.41$ & $1757.44$ & $7.334$ \\
PI-SSD & $8.56$ & $198.94$ & $0.044$ \\
SEGSID & $22.95$ & $2044.56$ & $112.511$ \\
\midrule
Proposed & $8.56$ & $198.94$ & $0.043$ \\
\bottomrule
\end{tabular*}
\end{table}

\section{Conclusion}
\label{sec:conclusion}

This work formulated multibeam imaging sonar restoration as a physics-consistent self-supervised inverse problem for joint despeckling and deblurring. Starting from a coherent stochastic observation model, we derived the conditional-mean intensity for the forward model that links latent reflectivity to detected sonar intensity. This model enables paired physics-consistency learning from registered multi-frame observations without clean reflectivity targets. To regularize the ill-conditioned inverse problem, we further introduced robust equivariant imaging based on sonar-specific forward simulation.

Experiments on the Boat and Underwater Structure datasets show that the proposed method improves paired forward prediction while achieving a favourable balance between speckle suppression and edge preservation. Compared with representative model-based and learning-based baselines, the proposed framework produces reconstructions that are more consistent with the physical forward model and less prone to either residual speckle or excessive structural smoothing.

The current framework relies on registered paired observations and an approximate local PSF, and may be affected by severe registration errors, dynamic scene changes, or forward-model mismatch. A remaining limitation is that real-sonar deblurring is evaluated indirectly, since clean latent reflectivity ground truth is unavailable. Future work will focus on controlled simulations or calibration-target experiments to directly quantify reflectivity accuracy and PSF-mismatch robustness, as well as broader deployment across sonar platforms, imaging geometries, and more diverse underwater environments.

\bibliographystyle{IEEEtran}
\bibliography{ref}

\ifCLASSOPTIONcaptionsoff
  \newpage
\fi

\end{document}


\begin{center}
    {\Large \textbf{Supplementary Material}}\\[0.6em]
    {\large Self-Supervised Multibeam Sonar Restoration via Coherent Forward Modeling}\\[0.6em]
    {\normalsize Zixin~Wang, James~Hopgood, Mike~Davies}
\end{center}

\section{Image registration Validation}
\label{sup:pair_diagnostics}

This section supports the paired-observation assumption used in the main paper. The image registration example illustrates the alignment of scene structures before paired training and PFPE evaluation, while the frame-gap analysis examines how temporal separation reduces input-target correlation.

\begin{figure}[!htbp]
    \centering
    \begin{subfigure}[t]{0.45\columnwidth}
        \centering
        \includegraphics[width=\linewidth]{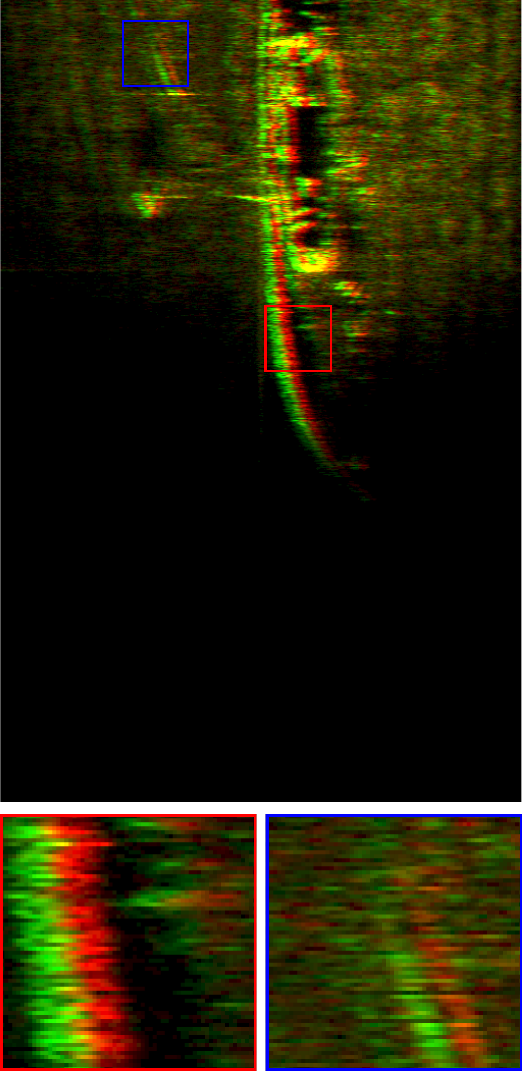}
        \caption{Before registration}
    \end{subfigure}
    \hfill
    \begin{subfigure}[t]{0.45\columnwidth}
        \centering
        \includegraphics[width=\linewidth]{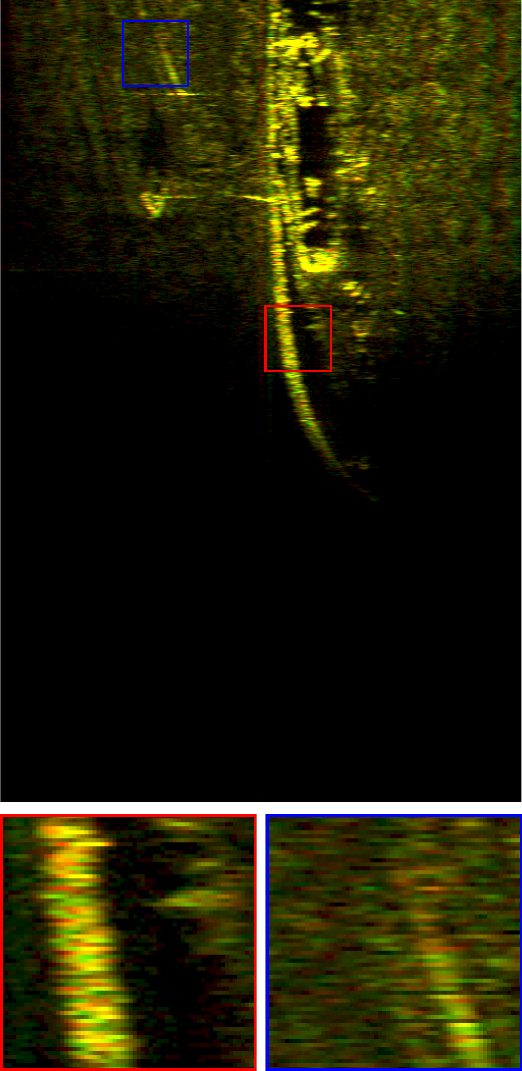}
        \caption{After registration}
    \end{subfigure}
    \caption{
Representative registration check on the Boat dataset. The reference frame $y_1$ is shown in red and the paired frame $y_2$ is shown in green; yellow indicates spatial overlap between the two frames. Registration increases the overlap of persistent scene structures, supporting the use of the pair for self-supervised training and PFPE evaluation.
}
    \label{fig:registration_check}
\end{figure}

Registered frame pairs are used for paired training and PFPE evaluation, whereas inference uses a single input frame. Candidate pairs are aligned by phase correlation following the paired observation registration described in the main paper. Fig.~\ref{fig:registration_check} shows a representative Boat example with an estimated translation of $\Delta \rho_1=9$ pixels and $\Delta \rho_2=0$ pixels. The reference frame $y_1$ is shown in red and the paired frame $y_2$ is shown in green; regions that overlap after alignment appear yellow. The increased yellow overlap after registration indicates improved alignment of persistent scene structures.

\section{Frame-Pair Correlation Analysis}
\label{sec:supp_frame_pair_correlation}

This section evaluates whether the registered frame pairs used for self-supervised training and PFPE evaluation provide a practical balance between reduced inter-frame correlation and reliable registration. Pairs that are too highly correlated may contain nearly identical speckle fluctuations, whereas pairs with too large a temporal separation may suffer from scene mismatch, viewpoint change, and registration uncertainty.

\paragraph{Correlation diagnostic}
For each frame gap \(\ell\), we compute the Pearson correlation coefficient \(r\) between registered frame pairs within homogeneous ROIs. These ROIs are selected to avoid obvious edges and structural boundaries, following the same principle used for ENL evaluation. Table~\ref{tab:gap_roi_correlation} reports the percentage of valid registered pairs below a practical reduced-correlation threshold of \(r<0.5\) as a function of frame gap. For each dataset, we select the smallest gap for which this percentage is at least $50\%$, subject to the displacement bounds used for registration. This rule helps avoid near-duplicate training pairs while limiting the viewpoint variation and registration uncertainty that arise at larger frame gaps. The threshold is not intended to prove statistical independence or pure speckle decorrelation, since real sonar ROIs may still contain residual texture, weak unresolved structure, and registration error.

\begin{table}[t]
\centering
\caption{Empirical ROI Pearson correlation between registered frame pairs at different frame gaps. The reduced-correlation ratio denotes the percentage of valid registered pairs with $r<0.5$.}
\label{tab:gap_roi_correlation}
\begin{tabular}{c cc cc}
\toprule
\multirow{2}{*}{Frame gap} &
\multicolumn{2}{c}{Boat} &
\multicolumn{2}{c}{Underwater Structure} \\
\cmidrule(lr){2-3}\cmidrule(lr){4-5}
& Pearson $r$  & $r<0.5$ 
& Pearson $r$  & $r<0.5$  \\
\midrule
1  & $0.485 \pm 0.138$ & $56.59\%$ & $0.269 \pm 0.174$ & $88.98\%$ \\
2  & $0.370 \pm 0.151$ & $79.29\%$ & $0.191 \pm 0.121$ & $96.71\%$ \\
3  & $0.294 \pm 0.156$ & $87.46\%$ & $0.156 \pm 0.100$ & $98.90\%$ \\
4  & $0.240 \pm 0.157$ & $94.15\%$ & $0.133 \pm 0.089$ & $99.61\%$ \\
5  & $0.208 \pm 0.153$ & $96.87\%$ & $0.118 \pm 0.083$ & $99.92\%$ \\
10 & $0.133 \pm 0.135$ & $98.91\%$ & $0.081 \pm 0.066$ & $100.00\%$ \\
\bottomrule
\end{tabular}
\end{table}

\paragraph{Displacement and overlap diagnostics}
Since reduced inter-frame correlation does not by itself guarantee reliable paired observations, we also examine geometric reliability after registration. Increasing the frame gap can reduce residual correlation, but it can also increase apparent motion, reduce the common valid image support, and introduce scene mismatch. We therefore report displacement and overlap diagnostics in addition to the ROI correlation analysis.

For each frame gap \(\ell\), phase correlation estimates the translational displacement
\((\Delta\rho_{1,\ell},\Delta\rho_{2,\ell})\)
between the input frame and its registered paired observation. We summarize the typical motion using the median absolute displacement
\begin{equation}
\mathrm{Med}_{\Delta\rho}(\ell)
=
\left(
\mathrm{median}\left(|\Delta\rho_{1,\ell}|\right),
\mathrm{median}\left(|\Delta\rho_{2,\ell}|\right)
\right),
\end{equation}
which is less sensitive than the mean to occasional large-motion or imperfect-registration pairs.

We also report the percentage of candidate pairs satisfying the fixed small-displacement condition
\begin{equation}
P_s(\ell)
=
100\,\Pr\left(
|\Delta\rho_{1,\ell}| \leq 10,\,
|\Delta\rho_{2,\ell}| \leq 5
\right),
\end{equation}
where the probability is computed over candidate registered pairs. A lower \(P_s(\ell)\) indicates that a larger fraction of frame pairs require substantial translation, which increases the risk of registration uncertainty and local scene mismatch.

To measure how much image support remains valid after translation, we report the percentage of high-overlap pairs:
\begin{equation}
P_o(\ell)
=
100\,\Pr\left(
\frac{
\left(H-|\Delta\rho_{2,\ell}|\right)
\left(W-|\Delta\rho_{1,\ell}|\right)
}{
HW
}
>0.95
\right),
\end{equation}
where \(H\) and \(W\) denote the image height and width. A lower \(P_o(\ell)\) indicates that larger frame gaps more frequently produce pairs with reduced valid support after registration.

Table~\ref{tab:frame_pair_displacement_overlap} shows that frame-gap selection involves a trade-off: larger gaps can reduce residual correlation, but they also make the paired observation geometrically less reliable. For this reason, all main cross-method comparisons use the same fixed evaluation gap, while larger-gap results are reported only as sensitivity analysis.

\begin{table}[t]
\centering
\caption{
Frame-pair displacement and overlap diagnostics for the Boat and Underwater Structure datasets.
$\mathrm{Med}_{\Delta \rho}$ denotes the median absolute displacement pair
$(\mathrm{median}|\Delta \rho_1|,\mathrm{median}|\Delta \rho_2|)$.
$P_s$ denotes the percentage of candidate pairs satisfying
$|\Delta \rho_1| \leq 10$ and $|\Delta \rho_2| \leq 5$.
$P_o$ denotes the percentage of candidate pairs with valid-overlap ratio
$q_{i,\ell}>0.95$ after translation.
}
\label{tab:frame_pair_displacement_overlap}
\begin{tabular}{ccccccc}
\toprule
\multirow{2}{*}{Gap}
& \multicolumn{3}{c}{Boat}
& \multicolumn{3}{c}{Underwater Structure} \\
\cmidrule(lr){2-4}
\cmidrule(lr){5-7}
& $\mathrm{Med}_{\Delta \rho}$
& $P_s(\%)$
& $P_o(\%)$
& $\mathrm{Med}_{\Delta \rho}$
& $P_s(\%)$
& $P_o(\%)$ \\
\midrule
1  & $(3,0)$  & 91.78 & 99.17 & $(3,0)$ & 88.83 & 99.84 \\
2  & $(5,0)$  & 68.24 & 94.48 & $(5,0)$ & 69.56 & 93.82 \\
3  & $(8,1)$  & 47.73 & 83.89 & $(6,0)$ & 60.66 & 85.58 \\
4  & $(10,1)$ & 38.89 & 72.35 & $(6,0)$ & 56.75 & 80.85 \\
5  & $(14,1)$ & 30.17 & 64.03 & $(6,0)$ & 54.25 & 75.94 \\
10 & $(28,2)$ & 17.00 & 39.37 & $(4,1)$ & 42.87 & 69.02 \\
\bottomrule
\end{tabular}
\end{table}

\paragraph{Physical Interpretation of Correlation Decay}
The two datasets exhibit different empirical correlation decay rates, even under the same frame-gap protocol. This difference should not be interpreted as a pure speckle statistic, because the measured ROI correlation is affected by platform motion, imaging range, scene geometry, viewpoint change, registration residuals, and weak unresolved structure within the selected homogeneous ROIs. Among the available acquisition parameters, imaging range provides useful physical context. The Boat dataset was acquired with a maximum range of $50~\mathrm{m}$, whereas the Underwater Structure dataset used a shorter maximum range of $25.5~\mathrm{m}$. For a lateral platform displacement $\Delta s$ at range $R$, the angular change scales approximately as $\Delta\theta\approx\Delta s/R$. Thus, for the same physical displacement, closer-range observations can undergo larger viewpoint changes, which alter the relative path lengths and phases of unresolved scatterers within a resolution cell. This can lead to stronger speckle decorrelation after registration. This physical interpretation offers one possible explanation for why the empirical correlation diagnostic is dataset dependent and supports using it only as a practical check of paired-frame suitability, rather than as proof of statistical independence.

\section{Results on the Boat Dataset}
\label{sup:boat_additional_results}

\begin{figure*}[t]
    \centering
    \captionsetup[subfigure]{font=footnotesize}
    \newcommand{\boatfigw}{0.18\textwidth}

    \begin{subfigure}[t]{\boatfigw}
        \centering
        \includegraphics[width=\linewidth]{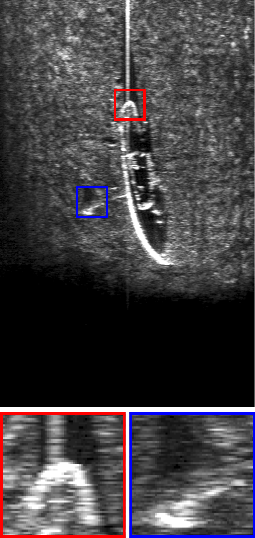}
        \caption{Noisy}
    \end{subfigure}
    \hfill
    \begin{subfigure}[t]{\boatfigw}
        \centering
        \includegraphics[width=\linewidth]{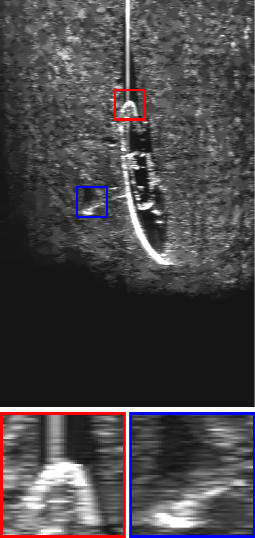}
        \caption{PPB}
    \end{subfigure}
    \hfill
    \begin{subfigure}[t]{\boatfigw}
        \centering
        \includegraphics[width=\linewidth]{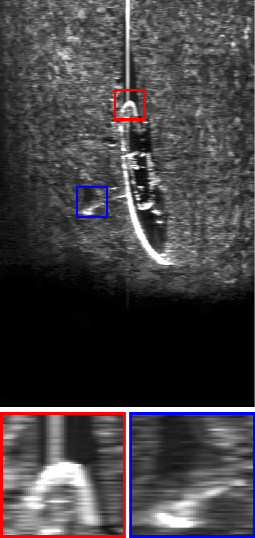}
        \caption{SAR-BM3D}
    \end{subfigure}
    \hfill
    \begin{subfigure}[t]{\boatfigw}
        \centering
        \includegraphics[width=\linewidth]{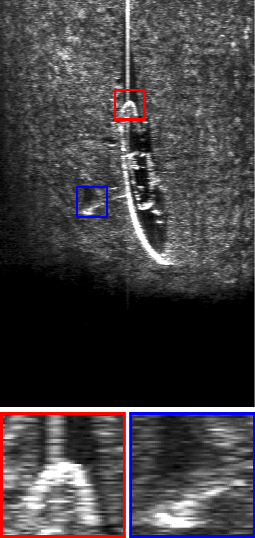}
        \caption{Blind2Unblind}
    \end{subfigure}
    \hfill
    \begin{subfigure}[t]{\boatfigw}
        \centering
        \includegraphics[width=\linewidth]{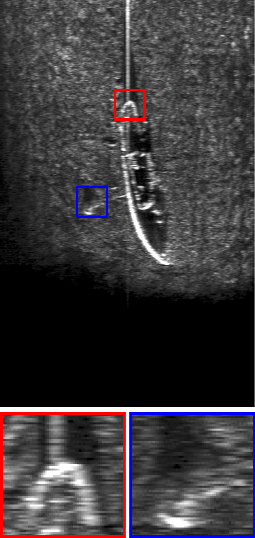}
        \caption{Noise2Void}
    \end{subfigure}

    \vspace{0.6em}

    \begin{subfigure}[t]{\boatfigw}
        \centering
        \includegraphics[width=\linewidth]{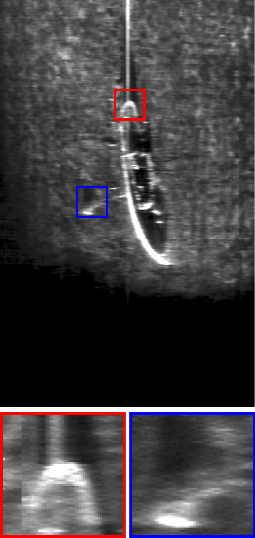}
        \caption{LG-BPN}
    \end{subfigure}
    \hfill
    \begin{subfigure}[t]{\boatfigw}
        \centering
        \includegraphics[width=\linewidth]{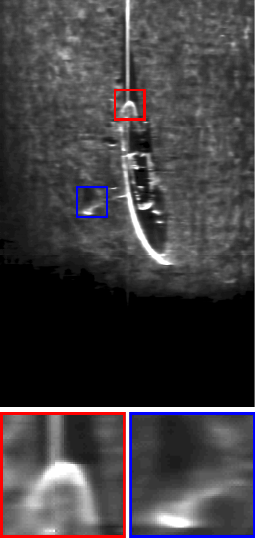}
        \caption{SS-BSN}
    \end{subfigure}
    \hfill
    \begin{subfigure}[t]{\boatfigw}
        \centering
        \includegraphics[width=\linewidth]{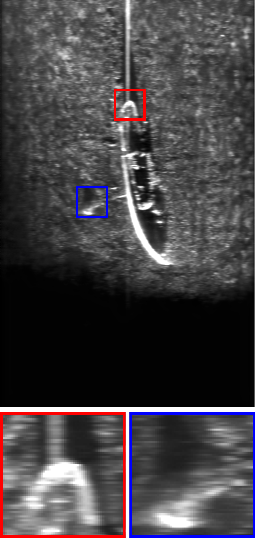}
        \caption{PI-SSD}
    \end{subfigure}
    \hfill
    \begin{subfigure}[t]{\boatfigw}
        \centering
        \includegraphics[width=\linewidth]{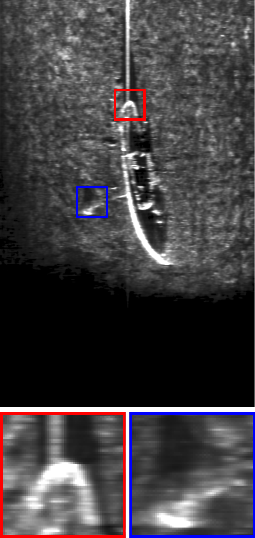}
        \caption{SEGSID}
    \end{subfigure}
    \hfill
    \begin{subfigure}[t]{\boatfigw}
        \centering
        \includegraphics[width=\linewidth]{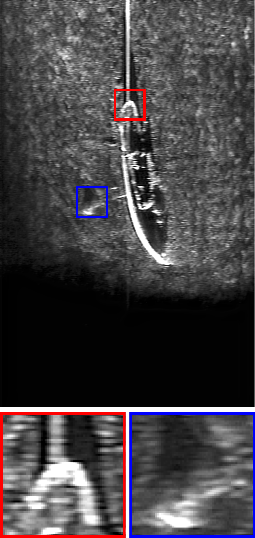}
        
        \caption{Proposed}
    \end{subfigure}
    \caption{
Additional qualitative comparison on the Boat dataset. Red and blue boxes indicate zoomed regions containing high-reflectivity structures and background-shadow transitions. The proposed method suppresses speckle while preserving separable local structures and clearer transition boundaries. This visual behavior is consistent with its higher PFPE improvement and lower $D_{\mathrm{EPI}}$ in Table~\ref{tab:sup_boat_full_results}.
}
    \label{fig:sup_boat_visual_results_2}
\end{figure*}

This section provides the complete Boat results supporting the main-paper comparison. Table~\ref{tab:sup_boat_full_results} reports the raw PFPE values, $\mathrm{PFPE}_{\mathrm{imp}}^{\mathrm{agg}}$,  ENL, EPI, and $D_{\mathrm{EPI}}$.
The additional metrics show why ENL alone is insufficient: several compared methods, such as SS-BSN and LG-BPN, increase ENL but also produce large $D_{\mathrm{EPI}}$, indicating that the apparent speckle reduction is associated with edge attenuation or oversmoothing rather than accurate forward prediction. In contrast, the proposed method achieves the highest $\mathrm{PFPE}_{\mathrm{imp}}^{\mathrm{agg}}$ while keeping EPI close to 1, supporting improved forward predictive agreement with limited deviation in observed edge
strength. Fig.~\ref{fig:sup_boat_visual_results_2} provides an additional qualitative example consistent with this trend.

\begin{table*}[t]
\centering
\caption{
Complete quantitative comparison on the Boat dataset. $\mathrm{PFPE}_{\mathrm{imp}}^{\mathrm{agg}}$ is computed from
the ratio of summed errors, which is relative to
the noisy paired-frame baseline. ENL measures homogeneous-region smoothing, while EPI and $D_{\mathrm{EPI}}$ characterize edge-strength preservation relative to the observed image. High ENL should therefore be interpreted together with PFPE and $D_{\mathrm{EPI}}$, since it can also result from oversmoothing.
}
\label{tab:sup_boat_full_results}
\begin{tabular*}{\textwidth}{@{\hspace{12pt}\extracolsep{\fill}} l l c c c c c @{\hspace{12pt}}}
\toprule
Category & Method 
& $\mathrm{PFPE}$  ($\times 10^{-4}$) $\downarrow$ 
& $\mathrm{PFPE}_{\mathrm{imp}}^{\mathrm{agg}}$ (\%) $\uparrow$ 
& ENL $\uparrow$ 
& EPI $\rightarrow 1$ 
& $D_{\mathrm{EPI}}$ (\%) $\downarrow$ \\
\midrule

Baseline 
& Noisy 
& $1.59 \pm 0.36$ 
& $0.00 $ 
& $8.18 \pm 1.12$ 
& $1.000 \pm 0.000$ 
& $0.00 \pm 0.00$ \\

\midrule
\multirow{2}{*}{Model-based}
& PPB 
& $1.79 \pm 0.26$ 
& $-13.06 $ 
& $12.73 \pm 2.71$ 
& $0.807 \pm 0.023$ 
& $19.32 \pm 2.28$ \\

& SAR-BM3D 
& $1.40 \pm 0.29$ 
& $ 11.53 $ 
& $12.14 \pm 2.40$ 
& $0.701 \pm 0.016$ 
& $29.87 \pm 1.60$ \\

\midrule
\multirow{6}{*}{Self-supervised}
& Noise2Void 
& $1.77 \pm 0.40$ 
& $-11.59 $ 
& $11.03 \pm 1.62$ 
& $0.734 \pm 0.014$ 
& $26.61 \pm 1.36$ \\

& Blind2Unblind 
& $1.37 \pm 0.29$ 
& $13.39 $ 
& $7.74 \pm 1.03$ 
& $1.047 \pm 0.004$ 
& $4.68 \pm 0.36$ \\

& LG-BPN 
& $ 1.57 \pm 0.31$
& $1.19 $ 
& $18.57 \pm 5.60$ 
& $0.541 \pm 0.023$ 
& $45.89 \pm 2.27$ \\

& SS-BSN 
& $1.62 \pm 0.32$  
& $ -2.24$ 
& $25.04 \pm 8.79$
& $0.441 \pm 0.031$ 
& $55.92 \pm 3.06$ \\

& PI-SSD 
& $1.44 \pm 0.27$
& $9.39$ 
& $16.97 \pm 4.81$
& $0.568 \pm 0.032$
& $43.22 \pm 3.24$ \\

& SEGSID 
& $1.46 \pm 0.31$ 
& $8.20$ 
& $15.83 \pm 3.81$ 
& $0.592 \pm 0.022$ 
& $40.82 \pm 2.16$ \\

\midrule
Proposed 
& Proposed
& \best{$1.36 \pm 0.28$}
& \best{$14.10$}
& $12.81 \pm 2.68$
& $0.992 \pm 0.062$
& $4.65 \pm 4.13$ \\

\bottomrule
\end{tabular*}
\end{table*}

\section{Results on the Underwater Structure Dataset}
\label{sup:underwater_structure_results}

\begin{figure*}[t]
    \centering
    \captionsetup[subfigure]{font=footnotesize}

    \begin{subfigure}[t]{0.19\textwidth}
        \centering
        \includegraphics[width=\linewidth]{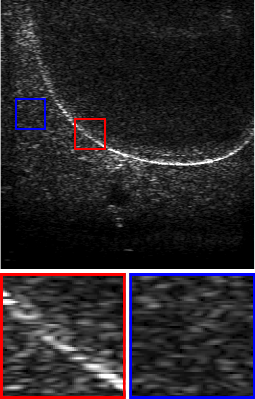}
        \caption{Noisy}
    \end{subfigure}
    \hfill
    \begin{subfigure}[t]{0.19\textwidth}
        \centering
        \includegraphics[width=\linewidth]{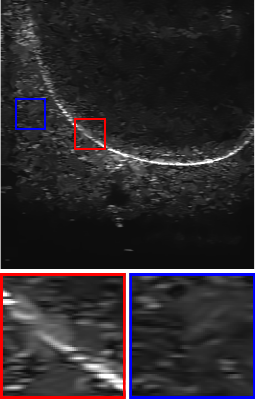}
        \caption{PPB}
    \end{subfigure}
    \hfill
    \begin{subfigure}[t]{0.19\textwidth}
        \centering
        \includegraphics[width=\linewidth]{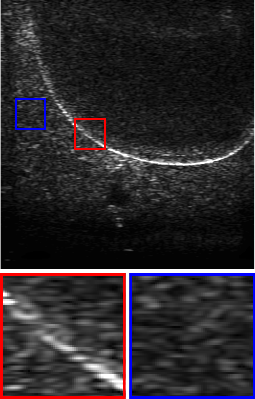}
        \caption{SAR-BM3D}
    \end{subfigure}
     \hfill
    \begin{subfigure}[t]{0.19\textwidth}
        \centering
        \includegraphics[width=\linewidth]{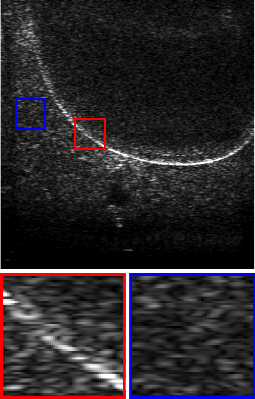}
        \caption{Blind2Unblind}
    \end{subfigure}
     \hfill
    \begin{subfigure}[t]{0.19\textwidth}
        \centering
        \includegraphics[width=\linewidth]{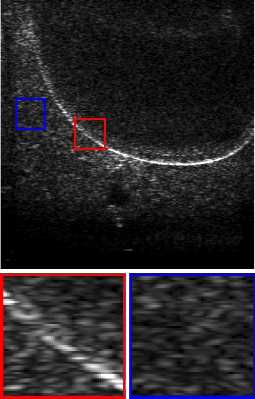}
        \caption{Noise2Void}
    \end{subfigure}
    \hfill

       \vspace{0.5em} 


    \begin{subfigure}[t]{0.19\textwidth}
        \centering
        \includegraphics[width=\linewidth]{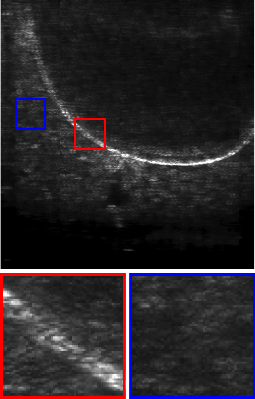}
        \caption{LG-BPN}
    \end{subfigure}
    \hfill
    \begin{subfigure}[t]{0.19\textwidth}
        \centering
        \includegraphics[width=\linewidth]{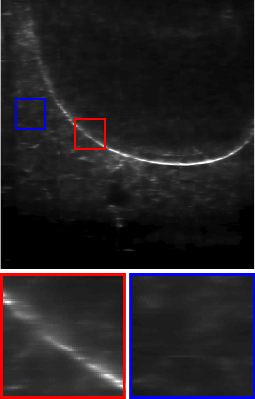}
        \caption{SS-BSN}
    \end{subfigure}
    \hfill
    \begin{subfigure}[t]{0.19\textwidth}
        \centering
        \includegraphics[width=\linewidth]{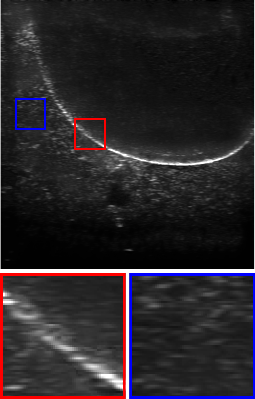}
        \caption{PI-SSD}
    \end{subfigure}
    \hfill
    \begin{subfigure}[t]{0.19\textwidth}
        \centering
        \includegraphics[width=\linewidth]{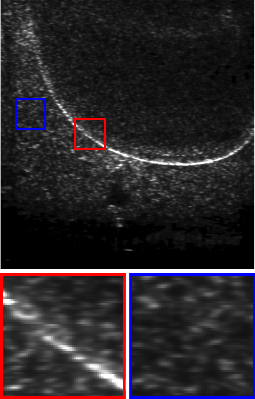}
        \caption{SEGSID}
    \end{subfigure}
    \hfill
    \begin{subfigure}[t]{0.19\textwidth}
        \centering
        \includegraphics[width=\linewidth]{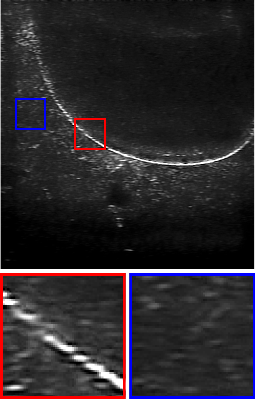}
        \caption{Proposed}
    \end{subfigure}
    \hfill
\caption{
Additional qualitative comparison on the Underwater Structure dataset. The zoomed regions highlight structural boundaries, weak internal reflectors, and background speckle. The proposed method suppresses speckle while preserving clearer structural continuity than representative baselines, consistent with its lower PFPE and relatively small $D_{\mathrm{EPI}}$ in Table~\ref{tab:sup_underwater_structure_results}.
}
\label{fig:sup_underwater_structure_visual}
\end{figure*}

This section provides the complete Underwater Structure results supporting the main-paper comparison. Table~\ref{tab:sup_underwater_structure_results} shows that the proposed method obtains the highest  $\mathrm{PFPE}_{\mathrm{imp}}^{\mathrm{agg}}$ while maintaining an EPI close to one. 
As in the Boat experiment, methods with substantially higher ENL, such as SS-BSN, also show much larger $D_{\mathrm{EPI}}$, indicating that stronger homogeneous-region smoothing does not necessarily imply better preservation of observed edge strength. Fig.~\ref{fig:sup_underwater_structure_visual} provides an additional qualitative comparison on a representative structural scene.

\begin{table*}[t]
\centering
\caption{
Complete quantitative comparison on the Underwater Structure dataset. $\mathrm{PFPE}_{\mathrm{imp}}^{\mathrm{agg}}$ is computed from the ratio of summed errors, which is relative to the noisy paired-frame baseline. ENL measures homogeneous-region smoothing, while EPI and $D_{\mathrm{EPI}}$ measure edge-strength preservation relative to the observed image. }
\label{tab:sup_underwater_structure_results}
\begin{tabular*}{\textwidth}{@{\hspace{12pt}\extracolsep{\fill}} l l c c c c c @{\hspace{12pt}}}
\toprule
Category & Method 
& $\mathrm{PFPE}$  ($\times 10^{-4}$) $\downarrow$ 
& $\mathrm{PFPE}_{\mathrm{imp}}^{\mathrm{agg}}$ (\%) $\uparrow$ 
& ENL $\uparrow$ 
& EPI $\rightarrow 1$ 
& $D_{\mathrm{EPI}}$ (\%) $\downarrow$ \\
\midrule

Baseline
& Noisy
& $6.72 \pm 3.06$
& $ 0.00$
& $3.38 \pm 0.36$
& $1.000 \pm 0.000$
& $0.00 \pm 0.00$ \\

\midrule
\multirow{2}{*}{Model-based}
& PPB
& $7.33 \pm 2.34$
& $ -9.01$
& $6.98 \pm 1.36$
& $0.728 \pm 0.042$
& $27.17 \pm 4.16$ \\

& SAR-BM3D
& $6.72 \pm 2.18$
& $ -0.02$
& $4.71 \pm 0.59$
& $0.746 \pm 0.011$
& $25.42 \pm 1.05$ \\

\midrule
\multirow{6}{*}{Self-supervised}
& Noise2Void
& $6.82 \pm 2.24$
& $ -1.49$
& $4.09 \pm 0.46$
& $0.828 \pm 0.006$
& $17.20 \pm 0.56$ \\

& Blind2Unblind
& $6.57 \pm 2.18$
& $2.25$
& $3.36 \pm 0.35$
& $1.011 \pm 0.001$
& $1.11 \pm 0.13$ \\

& LG-BPN
& $7.80 \pm 2.35$
& $-16.05$
& $7.11 \pm 1.35$
& $0.858 \pm 0.022$
& $14.19 \pm 2.16$ \\

& SS-BSN
& $8.85 \pm 2.84$
& $-31.66$
& $15.50 \pm 5.22$
& $0.288 \pm 0.039$
& $71.24 \pm 3.88$ \\

& PI-SSD
& $7.21 \pm 2.25$
& $-7.36$
& $7.76 \pm 2.02$
& $0.487 \pm 0.022$
& $51.29 \pm 2.19$ \\

& SEGSID 
& $6.86 \pm 2.22$
& $-2.02$
& $4.74 \pm 0.66$
& $0.711 \pm 0.014$
& $28.86 \pm 1.44$ \\

\midrule
Proposed
& Proposed
& \best{$6.43 \pm 2.08$}
& \best{$ 4.36 $}
& $6.37 \pm 2.84$
& $0.990 \pm 0.074$
& $ 6.83 \pm 2.95$ \\

\bottomrule
\end{tabular*}
\end{table*}

\section{Ablation and Sensitivity Analysis}
\label{sup:sensitivity_analysis}

This section reports additional Boat ablations and sensitivity analyses. Each block varies selected factors around the stated reference settings, so results are interpreted within each block rather than as main cross-method rankings.

Table~\ref{tab:sup_sensitivity} separates component ablations from hyperparameter sensitivity analyses, covering the paired physics-consistency loss, registration, REI transformation type, regularization weight, rotation magnitude, and number of simulated looks. We also analyze the effect of the training and evaluation frame gaps to distinguish restoration behaviour from the gap-dependent scaling of aggregate PFPE improvement. Since the Boat dataset exhibits more challenging scene variation and inter-frame displacement in our experiments, we use it for these additional analyses.

\subsection{Ablation Analysis}
The results show that the physics-consistency loss alone can improve the paired forward-prediction score, but does not produce a meaningful reflectivity reconstruction. Adding REI regularization is therefore necessary to mitigate the ill-conditioning left by the blurred forward operator. The physics-consistency-only ($\mathcal{L}_{\mathrm{PC}}$) variant obtains high PFPE improvement but collapses to a degenerate solution, with nearly zero ENL and an extremely large $D_{\mathrm{EPI}}$. This behaviour is consistent with the analysis in the main paper: the physics-consistency loss constrains $A f_\theta(y)$ in the measurement domain, but weakly determines high-frequency or weakly observed components of the latent reflectivity. The no-registration variant performs poorly in PFPE, confirming that paired supervision requires spatially aligned observations before the forward-prediction loss is meaningful.

\subsection{Sensitivity Analysis}

The sensitivity blocks in Table~\ref{tab:sup_sensitivity} show that the proposed method is governed by a trade-off between forward prediction, speckle suppression, and structural preservation. In each block, one factor is varied while the remaining settings are kept fixed, so the trends can be interpreted within each group.

For the transformation and rotation-angle blocks, $\alpha=0.05$ and $J=8$ are fixed; moderate geometric perturbations give the best balance, whereas very large rotations are less compatible with the locally shift-invariant PSF approximation and the anisotropic range-azimuth structure of sonar images.

For the REI-weight block, $J=8$ and the maximum rotation angle is fixed at $20^\circ$. Increasing $\alpha$ raises ENL, but overly strong regularization reduces PFPE improvement and attenuates edges, as indicated by lower EPI and larger $D_{\mathrm{EPI}}$. This confirms that higher homogeneous-region smoothness is not sufficient to indicate better restoration, since excessive regularization can lead to oversmoothing.

For the simulated-look block, $\alpha=0.05$ and the maximum rotation angle is fixed at $20^\circ$. The number of simulated looks $J$ controls the stochasticity of the REI forward simulation. Very small $J$ yields poorer PFPE and larger edge-strength deviation despite a higher ENL, indicating an unfavourable smoothing--preservation trade-off. As $J$ increases, the averaged simulator approaches the conditional-mean forward model; the observed PFPE improves and ENL decreases, while $D_{\mathrm{EPI}}$ remains similar for $J=8$ and $J=16$.

The reported setting is therefore used as a balanced operating point across PFPE, ENL, EPI, and $D_{\mathrm{EPI}}$, rather than being selected by optimizing any single metric.

\begin{table*}[t]
\centering
\caption{
Ablation and sensitivity analysis on the Boat dataset. 
$\mathrm{PFPE}_{\mathrm{imp}}^{\mathrm{agg}}$ is computed from the ratio of summed errors.
ENL measures homogeneous-region smoothing, while \(D_{\mathrm{EPI}}\) measures mean per-image deviation from unity in relative edge strength.
In each block, one design factor is varied while the remaining settings are kept fixed whenever possible. 
The results are interpreted jointly in terms of paired-frame forward prediction, speckle suppression, and structural preservation.}
\label{tab:sup_sensitivity}
\begin{tabular*}{\textwidth}{@{\hspace{12pt}\extracolsep{\fill}} l c c c c c @{\hspace{12pt}}}
\toprule
Variant
& $\mathrm{PFPE}$ ($\times 10^{-4}$) $\downarrow$
& $\mathrm{PFPE}_{\mathrm{imp}}^{\mathrm{agg}}$ (\%) $\uparrow$
& ENL $\uparrow$
& EPI $\rightarrow 1$
& $D_{\mathrm{EPI}}$ (\%) $\downarrow$ \\
\midrule

\multicolumn{6}{l}{\textit{Loss and registration}} \\
$\mathcal{L}_{\mathrm{PC}}$-only
& $1.24 \pm 0.29$
& $22.02$
& $0.02 \pm 0.00$
& $23.093 \pm 1.419$
& $2209.35 \pm 141.89$ \\

No registration
& $1.62 \pm 0.29$
& $-2.32$
& $14.88 \pm 3.80$
& $0.549 \pm 0.026$
& $45.09 \pm 2.59$ \\
\midrule
\multicolumn{6}{l}{\textit{Transformation}} \\
Scale
& $1.38 \pm 0.28$
& $13.07$
& $12.74 \pm 2.53$
& $0.956 \pm 0.054$
& $5.69 \pm 3.96$ \\

Shift
& $1.40 \pm 0.30$
& $11.97$
& $14.29 \pm 3.47$
& $0.926 \pm 0.076$
& $8.79 \pm 5.95$ \\

CPAB
& $1.41 \pm 0.28$
& $10.83$
& $13.58 \pm 3.23$
& $1.013 \pm 0.050$
& $ 3.92 \pm 3.33$ \\

Rot. + scale
& $1.42 \pm 0.29$
& $10.54$
& $14.77 \pm 3.24$
& $0.926 \pm 0.068$
& $8.23 \pm 5.75$ \\

Rot. + shift
& $1.49 \pm 0.31$
& $6.10$
& $15.19 \pm 3.98$
& $0.778 \pm 0.055$
& $22.23 \pm 5.50$ \\

\midrule
\multicolumn{6}{l}{\textit{Rotation angle}} \\
$10^\circ$
& $1.37 \pm 0.29$
& $13.88$
& $11.35 \pm 2.12$
& $0.944 \pm 0.043$
& $6.18 \pm 3.46$  \\

$20^\circ$ 
& $1.37 \pm 0.30$
& $13.37$
& $12.40 \pm 2.73$
& $0.968 \pm 0.058$
& $5.17 \pm 4.04$ \\

$45^\circ$
& $1.42 \pm 0.30$
& $10.56$
& $10.81 \pm 2.06$
& $0.822 \pm 0.039$
& $17.78 \pm 3.86$ \\

$90^\circ$
& $1.40 \pm 0.31$
& $11.86$
& $10.24 \pm 1.76$
& $0.887\pm 0.042$
& $11.35 \pm 4.21$ \\

\midrule
\multicolumn{6}{l}{\textit{REI weight $\alpha$}} \\
$\alpha=0.03$
& $1.35 \pm 0.29$
& $14.97$
& $9.98 \pm 1.70$
& $1.079 \pm 0.077$
& $8.56 \pm 6.97$ \\

$\alpha=0.05$
& $1.37 \pm 0.30$
& $13.37$
& $12.40 \pm 2.73$
& $0.968 \pm 0.058$
& $5.17 \pm 4.04$ \\

$\alpha=0.10$
& $1.58 \pm 0.32$
& $0.06$
& $17.37 \pm 5.81$
& $0.718 \pm 0.057$
& $28.22 \pm 5.71$ \\

$\alpha=0.50$
& $1.65 \pm 0.30$
& $-3.78$
& $26.57 \pm 11.38$
& $0.547 \pm 0.045$
& $45.30 \pm 4.53$ \\

\midrule
\multicolumn{6}{l}{\textit{Simulated looks $J$}} \\

$J=1$
& $1.83 \pm 0.32$
& $-15.13$
& $20.32 \pm 7.09$
& $0.500 \pm 0.031$
& $49.98 \pm 3.14$ \\

$J=4$
& $1.70 \pm 0.33$
& $-7.22$
& $19.10 \pm 6.55$
& $0.735 \pm 0.051$
& $26.49 \pm 5.13$ \\

$J=8$
& $1.37 \pm 0.30$
& $13.37$
& $12.40 \pm 2.73$
& $0.968 \pm 0.058$
& $5.17 \pm 4.04$ \\

$J=16$
& $1.34 \pm 0.29$
& $15.35$
& $12.10 \pm 2.42$
& $1.002 \pm 0.066$
& $5.24 \pm 4.06$ \\

\bottomrule
\end{tabular*}
\end{table*}

\subsection{Frame Gap Analysis}
\label{sec:supp_frame_gap_analysis}
This section analyzes how the training frame gap affects the restoration behaviour of the proposed method. All variants in Table~\ref{tab:boat_train_gap_test_gap} use the same configuration, with $\alpha=0.05$, $J=8$, and $T_\tau$ implemented as rotations with a maximum angle of $20^\circ$. For each training gap,  displacement tolerances are set to meet a 90\% pair-retention target to remove infeasible registration outliers, while the model and optimization settings are held fixed.

PFPE and $\mathrm{PFPE}_{\mathrm{imp}}^{\mathrm{agg}}$ are evaluated using fixed test gaps 1, 2, and 3, whereas ENL, EPI, and $D_{\mathrm{EPI}}$ are reported once for each trained checkpoint because they are computed from the restored single-frame output.

The $\mathrm{PFPE}_{\mathrm{imp}}^{\mathrm{agg}}$ is gap-dependent because it is normalized by the noisy paired-frame discrepancy. At a fixed test gap, identical test pairs, preprocessing, and evaluated image support yield a common noisy-pair denominator across methods; thus, aggregate normalization
preserves the ranking of mean PFPE.
Across test gaps, both the noisy baseline (the aggregate denominator changes with the paired-frame discrepancy at that gap) and the paired prediction error can change, so the scores are not directly comparable absolute image-quality measures. The main comparisons therefore use the common test gap 1, supported by the correlation and registration diagnostics in Section S2.

Table~\ref{tab:boat_train_gap_test_gap} shows this effect empirically. The noisy PFPE baseline increases from \(1.59\times10^{-4}\) at test gap 1 to \(2.35\times10^{-4}\) and \(2.94\times10^{-4}\) at test gaps 2 and 3, respectively. Under this pair-retention protocol, increasing the training gap from 1 to 2 and 3 increases ENL from \(12.40\) to \(15.31\) and \(16.48\), but also increases \(D_{\mathrm{EPI}}\) from \(5.17\%\) to \(15.82\%\) and \(20.91\%\). This indicates stronger smoothing but larger deviation from the observed edge strength.

The gap-1 trained model provides the most favourable balance in this analysis. It achieves the highest $\mathrm{PFPE}_{\mathrm{imp}}^{\mathrm{agg}}$ at test gaps 1 and 2 and the lowest $D_{\mathrm{EPI}}$. At test gap 3, the train-gap-3 model yields a slightly higher aggregate improvement (24.98\% versus 24.50\%), but a substantially larger $D_{\mathrm{EPI}}$ (20.91\% versus 5.17\%). Under this auxiliary configuration, these results support the choice of frame gap 1 for training in the main experiments.

\begin{table*}[t]
\centering
\caption{
Effect of the training frame gap on Boat test performance.
Model and optimization settings are fixed, while gap-specific displacement tolerances are set to meet a common 90\% pair-retention target.
PFPE is reported in units of $10^{-4}$, and Imp. denotes $\mathrm{PFPE}_{\mathrm{imp}}^{\mathrm{agg}}$ in percent.
PFPE and Imp. are evaluated using fixed PFPE test gaps 1, 2, and 3.
For each trained checkpoint, ENL, EPI, and $D_{\mathrm{EPI}}$ are reported once because they are computed from the restored single-frame output.
}
\label{tab:boat_train_gap_test_gap}
\begin{tabular*}{\textwidth}{@{\hspace{12pt}\extracolsep{\fill}} lccccccccc @{\hspace{12pt}}}
\toprule
&
\multicolumn{2}{c}{Test gap 1}
& \multicolumn{2}{c}{Test gap 2}
& \multicolumn{2}{c}{Test gap 3}
& & & \\
\cmidrule(lr){2-3}
\cmidrule(lr){4-5}
\cmidrule(lr){6-7}
Variant
& PFPE $\downarrow$
& Imp. $\uparrow$
& PFPE $\downarrow$
& Imp. $\uparrow$
& PFPE $\downarrow$
& Imp. $\uparrow$
& ENL $\uparrow$
& EPI $\rightarrow 1$
& $D_{\mathrm{EPI}}$ (\%) $\downarrow$ \\
\midrule
Noisy
& \(1.59 \pm 0.36\) & 0.00
& \(2.35 \pm 0.58\) & 0.00
& \(2.94 \pm 0.82\) & 0.00
& \(8.18 \pm 1.12\)
& \(1.000 \pm 0.000\)
& \(0.00 \pm 0.00\) \\

Train gap 1
& \(1.37 \pm 0.30\) & 13.37
& \(1.82 \pm 0.48\) & 22.20
& \(2.22 \pm 0.75\) & 24.50
& \(12.40 \pm 2.73\)
& \(0.968 \pm 0.058\)
& \(5.17 \pm 4.04\) \\

Train gap 2
& \(1.52 \pm 0.30\) & 4.04
& \(1.91 \pm 0.46\) & 18.58
& \(2.27 \pm 0.71\) & 22.60
& \(15.31 \pm 4.67\)
& \(0.842 \pm 0.066\)
& \(15.82 \pm 6.60\) \\

Train gap 3
& \(1.55 \pm 0.26\) & 2.30
& \(1.89 \pm 0.38\) & 19.45
& \(2.20 \pm 0.59\) & 24.98
& \(16.48 \pm 3.90\)
& \(0.791 \pm 0.058\)
& \(20.91 \pm 5.79\) \\
\bottomrule
\end{tabular*}
\end{table*}